**RESEARCH ARTICLE**

# A comparative study of convolutional neural network models for wind field downscaling

**Kevin Höhlein[1]**  | **Michael Kern[1]**  | **Timothy Hewson[2]**  | **Rüdiger Westermann[1]**

[1]TUM Department of Informatics, Technical University of Munich, Garching, Germany

[2]Forecast Department, European Center for Medium-Range Weather Forecasts, Reading, UK

**Correspondence**
Kevin Höhlein, TUM Department of Informatics, Technical University of Munich, Garching, DE-85748, Germany.
Email: kevin.hoehlein@tum.de

**Funding information**
Deutsche Forschungsgemeinschaft, Grant/Award Number: CRC/Transregio 165, Waves to Weather, Projects A7 and B5; Open access funding enabled and organized by Projekt DEAL

**Abstract**

We analyze the applicability of convolutional neural network (CNN) architectures for downscaling of short-range forecasts of near-surface winds on extended spatial domains. Short-range wind forecasts (at the 100 m level) from European Centre for Medium Range Weather Forecasts ERA5 reanalysis initial conditions at 31 km horizontal resolution are downscaled to mimic high resolution (HRES) (deterministic) short-range forecasts at 9 km resolution. We evaluate the downscaling quality of four exemplary CNN architectures and compare these against a multilinear regression model. We conduct a qualitative and quantitative comparison of model predictions and examine whether the predictive skill of CNNs can be enhanced by incorporating additional atmospheric variables, such as geopotential height and forecast surface roughness, or static high-resolution fields, like land–sea mask and topography. We further propose DeepRU, a novel U-Net-based CNN architecture, which is able to infer situation-dependent wind structures that cannot be reconstructed by other models. Inferring a target 9 km resolution wind field from the low-resolution input fields over the Alpine area takes less than 10 ms on our graphics processing unit target architecture, which compares favorably to an overhead in simulation time of minutes or hours between low- and high-resolution forecast simulations.

## 1 | INTRODUCTION AND CONTRIBUTION

Accurate prediction of near-surface wind fields is a topic of central interest in various fields of science and industry. Severe memory and performance costs of numerical weather simulations, however, limit the availability of fine-scale (high-resolution) predictions, especially when forecast data are required for extended spatial domains. While running global reanalyses and forecasts with a spatial resolution of around 30 km is computationally affordable (e.g., Hersbach *et al.*, 2020), these models are unable to reproduce wind climatology accurately in regions with complex orography, such as mountain ranges. Since wind speed and direction are determined by localized interactions between airflow and surface topography, with sometimes the added complication of thermal forcing, accurate numerical simulation requires information on significantly finer length scales, particularly in regions that are topographically complex. For instance, (sub-grid-scale) topographic features such as steep slopes, valleys, mountain ridges or cliffs may induce wind shear, turbulence, acceleration and deceleration patterns that cannot be resolved by global models that lack information on these factors. Moreover, meteorologically relevant factors such as the vertical stability, snow cover or the presence of nearby lakes, river beds or sea can strongly influence local wind conditions (e.g., McQueen et al., 1995; Holtslag et al., 2013). In these regions, finer-resolution regional numerical models with grid spacings of the order of kilometers or less need to be applied in order to obtain reliable low-level winds (e.g., Salvador et al., 1999; Mass et al., 2002).

One approach to circumvent costly high-resolution simulations over extended spatial scales is known as downscaling, that is, inferring information on physical quantities at local scale from readily available low-resolution simulation data using suitable refinement processes. Downscaling is a long-standing topic of interest in many scientific disciplines, and in particular in meteorological research there exists a large variety of methods to downscale physical parameters. Such methods can be broadly classified into dynamical and empirical-statistical approaches (e.g., Hewitson and Crane, 1996; Rummukainen, 1997; Wilby and Wigley, 1997).

In dynamical downscaling (e.g., Räisänen et al., 2004; Rummukainen, 2010; Radić and Clarke, 2011; Kotlarski et al., 2014; Xue et al., 2014), high-resolution numerical models are used over limited sub-domains of the area of interest, and numerical model outputs on coarser scales provide boundary conditions for the simulations on a finer scale. While the restricted size of the model domain leads to a significant reduction of computational costs compared to global domain simulations, dynamical downscaling still remains computationally demanding and time-consuming.

Statistical downscaling, on the other hand, aims to avoid simulation at the finer scales by using a coarse-scale simulation (referred to as predictor data) to infer predictions at fine scale (referred to as predictand data). Correlations between the quantities at fine and coarse scales are learned by training statistical models on a set of known predictor–predictand data pairs.

Over time, a large number of empirical-statistical downscaling approaches have been developed, which apply statistical regression methods for downscaling purposes, such as (generalized) multilinear regression methods (e.g., Chandler, 2005) or quantile mapping approaches (e.g., Wood et al., 2004). With recent developments in data-driven machine learning and computer science, however, more powerful modeling techniques have become available, which may have the potential to outperform previous methods in terms of both accuracy and efficiency. Only a few studies have examined the use of nonlinear regression methods or more recent non-classical machine learning techniques (e.g., Eccel et al., 2007; Gaitan et al., 2014; Vandal et al., 2019). Specifically, the extent to which nonlinear machine learning approaches can provide additional value over classical methods is a question that has not been answered conclusively, as yet.

Deep learning methods are among the most prominent examples of state-of-the-art machine learning techniques (e.g., LeCun et al., 2015; Goodfellow et al., 2016). In particular, convolutional neural networks (CNNs) have found manifold applications in complex image processing and understanding tasks (e.g., Guo et al., 2016; Yang et al., 2019). One of these is single-image super-resolution, that is, the generation of high-resolution images from low-resolution images (e.g., Yang et al., 2019), which, formally, can be thought of as a very similar task to downscaling of climate variables.

CNNs rely on expressing regression models that operate on an extended spatial domain as a set of localized linear models (localized filter kernels), which are applied repeatedly at varying spatial positions across the domain through convolution operations. The restriction of the model parametrization to local filter kernels effectively limits the number of trainable parameters, and thus reduces the tendency of the model to overfit spurious patterns in the data, while increasing model efficiency. While also applicable to irregular graph-based data structures (Kipf and Welling, 2016), for example data defined on irregular grids, CNNs work most effectively with regular-gridded data in multi-dimensional array representations, facilitating an efficient parallel computation of optimization tasks on graphics processing unit (GPU) based computer hardware. Computational efficiency through parallelization is one of the major selling points of CNNs and should be considered as an important aspect during model design and data preparation. Furthermore, more complex mappings can be learned by stacking multiple layers of convolution operations (increasing the depth of the models) and applying these successively to generate more abstract feature representations. Similar to standard artificial neural networks, applying nonlinear activation functions between successive convolution layers can enable the model to learn nonlinear mappings. Beyond purely sequential feature processing, more elaborate model design patterns, like skip connections between pairs of convolution layers (Srivastava et al., 2015), residual learning (e.g., He et al., 2016) or changes in the spatial resolution of internal feature representations (e.g., Ronneberger et al., 2015), can be leveraged to improve model performance.

CNNs are thus particularly well suited for learning tasks involving spatially distributed data, which are often encountered in meteorology. Although CNN-based model architectures are increasingly adopted also in Earth-system sciences (e.g., Shen, 2018; Reichstein et al., 2019; Vannitsem et al., 2020), their use for downscaling applications has rarely been discussed (e.g., Vandal et al., 2018; Baño-Medina et al., 2019). In particular, earlier studies focused on simple CNN architectures which do not make use of recent model design patterns and thus do not exploit the full potential of state-of-the-art CNN architectures.

## 1.1 | Contribution

In this work, we perform a study of fully-convolutional neural network architectures for statistical downscaling of near-surface wind vector fields. The results are compared to those obtained by a multilinear regression model, with respect to both quality and performance. We train models to predict the most likely outcome of a high-resolution simulation of near-surface winds 100 m above ground, based on low-resolution short-range wind field forecasts as primary predictors. The data are defined on irregular octahedral and triangular reduced Gaussian grids with 9 km and 31 km horizontal resolution, respectively. To enable efficient processing of the data with CNNs and to avoid destroying local detail via interpolation, the data are mapped to regular grids through suitable padding. We view this work as an initial "proof of concept" step, to pave the way to using finer resolutions, for both predictor and predictand. If the predictand scale could reach 1 or 2 km we would envisage a much greater range of practical applications emerging.

We compare the capabilities of different existing models, which reflect varying degrees of model complexity and elaboration. Starting with a multilinear regression model and a light-weight linear convolutional model, we continue the comparison with nonlinear convolutional models of increasing complexity. By incorporating beneficial design patterns identified beforehand, in combination with adaptations in architectural design and training methodology, we propose DeepRU—a U-Net-based CNN model that improves the reconstruction quality of existing architectures.

For all models, we analyze whether incorporating additional climate variables and high-resolution topography like surface altitude and land–sea mask (LSM) improves the network's inference capabilities. We further train the models on sub-regions of the domain, to avoid learning relationships between low- and high-resolution winds based on geographical location, that is, to avoid overfitting to a particular domain. The reconstruction quality of all downscaling models is compared to the high-resolution simulations of real-world weather situations for a topographically complex region in central and southern Europe for the period between March 2016 and September 2019 (Figure 1). Our key finding is that thought-out architecture design and appropriate model tuning enable network-based downscaling methods to generate high-resolution wind fields efficiently in which local- and global-scale structures are reproduced with high fidelity.

To further analyze the usability of network-based downscaling, the relationships between model complexity, network performance and computational requirements such as memory consumption and prediction time are evaluated. We show how the model depth as well as the design patterns used, that is, residual connections across successive convolution layers and U-shaped encoder–decoder architectures, are leveraged to balance between model complexity and prediction quality.

We have made our implementations publicly available at Höhlein and Kern (2020).

## 2 | RELATED WORK

## 2.1 | Empirical-statistical downscaling

In describing downscaling options available at the time, Wilby and Wigley (1997) distinguish between regression methods, weather typing approaches and stochastic weather generators. Regression-based methods build upon the construction of parametric models, which are trained in an optimization procedure to establish a transfer function between low-resolution predictor variables and high-resolution predictands. Weather typing approaches, in contrast, rely on finding a suitable match between a set of predictor values and predictor value sets contained in the training data, in order to select out the most appropriate weather pattern analogue (e.g., Zorita and von Storch, 1999). Stochastic weather generators provide a probabilistic approach and are trained to replicate spatio-temporal sample statistics, as implied by the training data (e.g., Wilks, 2010; 2012).

A comprehensive review and comparison of empirical-statistical models for downscaling climate variables has been conducted by Maraun et al. (2015; 2019) and Gutiérrez et al. (2019), who showed that many of the approaches perform well generally but leave space for improvement. For instance, realistic replication of spatial variability in the high-resolution predictand variables remains a major challenge for many of the models (Maraun et al., 2019).

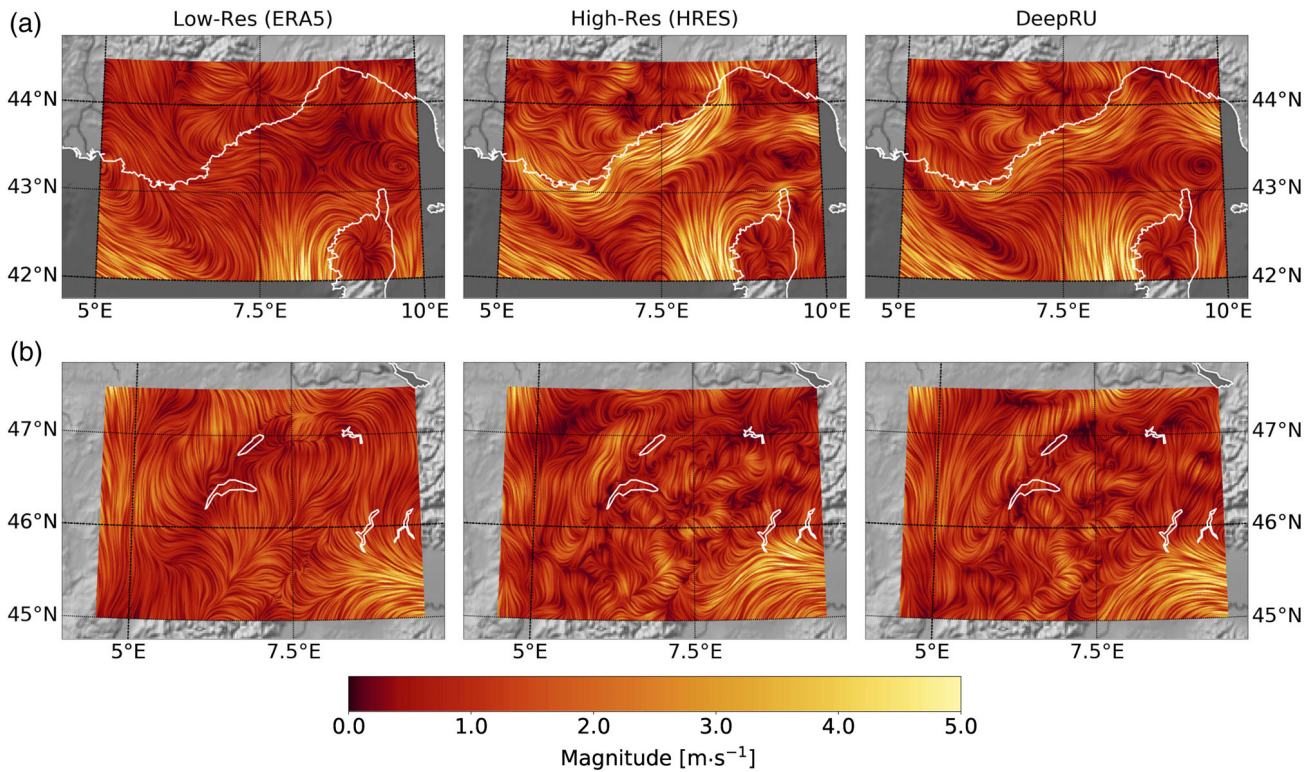

**FIGURE 1** Wind field on December 5, 2018, at 1200 UTC. Left: Low-resolution simulation based on ERA5 reanalysis data. Middle: High-resolution simulation based on HRES. Right: Prediction from the low-resolution field, our proposed convolutional neural network DeepRU. Streamlines are color coded with wind magnitude. (a) Coastal region enclosing the French Riviera and Corsica. (b) Highly varying winds over part of the Swiss Alps

Specifically addressing the problem of wind field downscaling and forecasting, Pryor (2005) and Michelangeli et al. (2009) proposed distribution-based approaches for wind field inference, and Huang et al. (2015) proposed a physical-statistical hybrid method for downscaling.

The question of what methods provide additional value over classical approaches has only been addressed by a number of smaller model comparison studies—with varying results. While Eccel et al. (2007), Mao and Monahan (2018) and Vandal et al. (2019) found hardly any or no advantage in applying non-classical machine learning methods, Gaitan et al. (2014) show non-classical methods outperforming classical ones, with artificial neural networks being a particular method example. More recently, Buzzi et al. (2019) used neural networks for nowcasting wind in the Swiss Alps and achieved very skillful models. These apparently contradictory findings raise the question of when, and under what conditions, deep learning methods can be profitably employed for downscaling.

Within meteorology, only a small number of studies have dealt with using CNNs for downscaling applications.

For example, Vandal et al. (2018) proposed "DeepSD," a simple CNN for downscaling precipitation over extended spatial domains, and more recently Baño-Medina et al. (2019) studied the performance of a set of CNNs for downscaling temperature and precipitation over Europe. Pan et al. (2019) proposed a similar architecture, again with a focus on precipitation.

While the influence of model complexity has been examined by Baño-Medina et al. (2019) in terms of model depth, that is, the number of convolution layers, the models in use did not exploit recent design patterns, like skip or residual connections (e.g., Srivastava et al., 2015; He et al., 2016) or the fully-convolutional U-Net-like architecture (Ronneberger et al., 2015), which enable network models to achieve state-of-the-art results in computer vision tasks.

## 2.2 | Upscaling of images and physical fields

Computer vision, being the origin of a large number of technological developments in machine learning, provides a problem setting, which is closely related to downscaling in meteorology and climatology—single-image super-resolution. There, the goal is to identify mappings

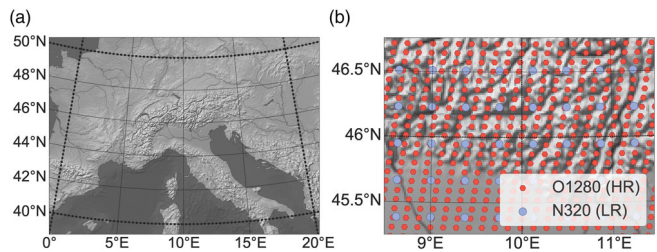

**FIGURE 2** (a) Map of the surface topography in Europe representing the data domain. (b) Low-resolution (N320) and high-resolution octahedral Gaussian simulation grid (O1280) used by ERA5 and HRES respectively. Over our domain the high-resolution grid comprises about three times more grid points in longitude and about four times more in latitude which allow the resolution of single low-resolution input images to be increased, while maintaining visual quality and avoiding pixel artifacts and blurriness. Within this context, the use of deep learning has led to remarkable improvements compared to standard statistical models (e.g., Yang *et al.*, 2019). In particular, CNNs were found to be particularly successful (e.g., Dong *et al.*, 2014; 2016a; 2016b; Sajjadi *et al.*, 2017).

Also in scientific data visualization researchers have begun to explore the capabilities of CNNs for upscaling and reconstruction of 2D/3D steady and time-varying scientific data, including both scalar and vector fields. Zhou *et al.* (2017) presented a CNN-based solution that downscales a volumetric dataset using three hidden layers designed for feature extraction, nonlinear mapping and reconstruction, respectively. Han *et al.* (2019) took a two-stage approach for vector field reconstruction via deep learning. The first stage initializes a low-resolution vector field based on the input streamline set. The second stage refines the low-resolution vector field to a high-resolution one via a CNN. The use of neural-network-based inference of data samples in the context of in situ visualization was demonstrated by Han and Wang (2020), by letting networks learn to infer missing time steps between 3D simulation results. Guo *et al.* (2020) designed a deep learning framework that produces spatial super-resolution of 3D vector field data. They demonstrate the downsampling of vector field data at simulation time and upsample the reduced data back to the original resolution. Weiss *et al.* (2019) extend image upscaling to geometry images of isosurfaces in 3D scalar fields by including depth and normal information.

# 3 | TRAINING DATA

For model training and evaluation, we use short-range weather forecast data, which include near-surface wind field simulations at different scales. The data are taken from the European Center for Medium Range Weather Forecasts (ECMWF) Meteorological Archival and Retrieval System (MARS) (Maass, 2019) and cover a spatial domain in central and southern Europe.

## 3.1 | Domain description

The training domain is restricted to 40 °–50 ° N and 0 °–20 ° E (Figure 2a) and is composed of sub-regions with varying orographic properties. Specifically, the domain contains high mountains of the Alps, some smaller mountain ranges in central Europe, flat areas in France, parts of the Mediterranean Sea and southwest-facing coastal regions of the Adriatic, to confront the employed models with challenging scenarios where winds are highly influenced by the topography. In particular, in the Dinaric Alps, situated in the eastern part of the domain, topographically forced gap flows are known to be an important phenomenon (e.g., Lee *et al.*, 2005; Belušić *et al.*, 2013). Significant differences between the low- and high-resolution numerical simulation results are most commonly observed in and around mountain ranges and coast lines, leading to the question of whether downscaling techniques can learn these differences and accurately predict the high-resolution fields from the low-resolution versions.

## 3.2 | Low- and high-resolution simulations

As "low-resolution" input to our models, we use data derived from the ERA5 reanalysis product suite (Hersbach *et al.*, 2020). ERA5 is the fifth in the series of ECMWF global reanalyses and provides estimates of the 3D global atmospheric state (climate) over time, based on a 4D variational data assimilation of past observations into a recent version of the operational ECMWF numerical forecast model. Output is provided on a regular reduced Gaussian grid with a horizontal resolution of 31 km ($0.28125°$). In this study we use hourly forecast fields, from data times of 0600 and 1800 UTC, at time steps of $T + 1, 2, ..., 12$ hr. We use these short-range forecasts instead of the true reanalysis fields to avoid systematic small jumps in low-level winds seen in the latter at 0900 and 2100 UTC (documented in Hersbach *et al.*, 2020).

The higher-resolution target dataset was provided by operational short-range forecasts from ECMWF's high-resolution (HRES) model, also at hourly intervals, initialized twice per day. HRES is a component of the ECMWF

Integrated Forecast System that can provide relatively accurate forecast products into the medium ranges ($\geq$72 hr ahead) (ECMWF, 2017). HRES is the highest available resolution model at ECMWF ($\sim$9 km) and, as with reanalyses, incorporates observations and information about the Earth-system as a prior for simulation runs. The output is provided on an octahedral reduced Gaussian grid (O1280). Forecast time steps used were $T + 7, 8, 9, ..., 18$ hr from the 0000 UTC and 1200 UTC runs. These were chosen as a compromise between being long enough to reduce any contamination from model spin-up and short enough to retain forecast accuracy. The different spatial resolutions of ERA5 and HRES are illustrated in Figure 2b.

Products for HRES on the O1280 grid were first introduced operationally in March 2016 and so are only available from that point onwards. Therefore, we restrict our analysis to time periods between March 2016 and October 2019.

## 3.3 | Predictor and predictand variables

Both the low-resolution predictors and the high-resolution predictands provide two wind variables, which contain spatio-temporal information on the horizontal wind components 100 m above ground. The wind variables are denoted by $U$ (meridional wind) and $V$ (zonal wind). At the same locations (i.e., grid points), land surface elevation (altitude, ALT) and a binary LSM are available in low- and high-resolution variants. These are used as static predictors.

From the low-resolution dataset, supplementary predictor variables are obtained and used as dynamical, that is, time-varying, predictors. The additional variables were manually selected according to the following considerations:

• Boundary layer height (BLH) is a model diagnostic that describes the vertical extent of the lowest layer of the atmosphere within which interactions take place between the Earth's surface and the atmosphere (Stull, 2017). Its value typically ranges between about 0.3 and 3 km and it is essentially a metric for low-level stability, with larger values implying deeper layers of instability-driven mixing. Earlier studies (e.g., Holtslag *et al.*, 2013) found that boundary-layer effects can have a significant impact on model performance in numerical temperature and wind predictions. Therefore, BLH may encode information that affects the matching between the low- and high-resolution variants. Also, BLH can provide the model with information about diurnal cycles. For these various reasons there was clear potential for this standard model output variable to be a useful predictor.

• Forecast surface roughness (FSR) denotes the surface roughness as represented in the forecast and thereby provides information on the frictional retardation of the near-surface airflow. Contributory factors are vegetation types and land cover such as soil or snow. The only dynamic component in the ECMWF modeling architecture is snow cover; other aspects are fixed year-round. We expected a small but direct impact from the snow cover.

• Geopotential height at 500 hPa (Z500) designates the elevation of the 500 hPa pressure level above mean sea level, and typically has values around 5,500 m. At this height, the pressure gradients and Coriolis force are typically in balance and winds are roughly parallel to Z500 isolines (see, for example, geostrophic winds in Wallace and Hobbs, 2006). Fields of Z500 very commonly serve as a proxy for forecasters of the general atmospheric flow structure and indeed synoptic pattern. So on the one hand one might expect a link with near-surface winds, but on the other the level is so far from the surface that it is unlikely to be a good predictor of local winds. This variable was partly included as a test of the veracity of our results. Even though on physical grounds we did not, overall, expect strong predictive skill from this variable, our results indicate an apparent influence on the inferred fields.

## 3.4 | Data padding

The training data obtained from MARS is defined on irregular grids where the number of grid nodes per latitude decreases with increasing latitude. As CNNs require the input data as multi-dimensional data arrays, the data need to be resampled on a regular grid structure. Since resampling using interpolation can smooth out and even remove relevant structures, the initial data are copied into rectangular 2D grids and padded appropriately. Therefore, the maximum number of longitudes for the latitude nearest to the equator is computed, and new points are padded to the remaining latitudes for each grid (cf. Figure 3). This approach preserves the spatial adjacency of grid nodes for a large proportion of the nodes, which is important to facilitate proper learning of spatial correlations. The true distance between grid nodes in world space is ignored, however, in the training process. The padded points are marked in a binary mask, which is passed to the objective function during network training to distinguish between valid and padded values in the loss computation.

**FIGURE 3**   Example of padding and masking used to resample the initial (low-resolution) data from an irregular Gaussian grid to a Cartesian grid. Blue cells indicate the data points of the gridded wind field. The interior of the data domain is shown in light-blue, boundary points are drawn in dark-blue, and their values are represented by numbers. A regular grid is achieved by padding new data points to the grid (light-red cells) while replicating the corresponding boundary values

Padding is chosen based on the fact that CNNs do not take into account only neighborhood relations but also relative changes of neighboring values. Zero padding, which may cause steep gradients between neighboring values, is thus deemed unsuitable and replaced by replication padding using the values of the boundary grid points of the valid domain.

The initial low- and high-resolution data with respectively 1,918 and 20,416 grid points on irregular grids are mapped to regular grids of size $36 \times 60$ and $144 \times 180$ in latitude and longitude directions. This results in an increase in the number of grid points by a factor of $4 \times 3$ between low-resolution and high-resolution grids, which reflects the actual difference in resolution between ERA5 and HRES simulations (see Figure 2).

## 3.5 | Data scaling

Before training, the padded data are standardized by subtracting sample mean and dividing by sample standard deviation. Standardization has proved useful in machine learning for improving the stability and convergence time of nonlinear optimization methods (e.g., Srivastava *et al.*, 2014; Ioffe and Szegedy, 2015). For time-dependent predictors, sample mean and standard deviation were computed node-wise from the snapshot statistics of the respective training datasets. Node-wise scaling is preferred over global domain scaling as spatial inhomogeneities are reduced, which we found to improve the downscaling results in our experiments. For static predictors, mean and standard deviation were computed from domain statistics. For sample standard deviations, we considered the unbiased ensemble estimate. Validation data are transformed accordingly before processing.

Standardization is performed also for the predictand variables. We found this useful due to strong differences in average wind speeds between coast or sea sites and mountain ranges. Further details are discussed in Section 5.3.

## 4 | NETWORK ARCHITECTURES

All of the models we use and compare in this work are constructed as parametric mappings of the form

$$y = f(x|\beta) \qquad (1)$$

where $y$ represents the array of high-resolution predictands, $x$ denotes the array of predictor variables and $\beta$ summarizes the model-specific parameters to be optimized during training. We use in particular CNNs, which repeatedly apply convolution kernels of fixed size to gridded input data at varying spatial positions to capture different types of features.

For the downscaling CNNs in our study, we consider input predictor arrays of shape $c_X^{(LR)} \times s_{lat} \times s_{lon}$ or $c_X^{(HR)} \times 4s_{lat} \times 3s_{lon}$, for low-resolution or high-resolution predictors $x^{(LR)}$ and $x^{(HR)}$ respectively. Here, $c_X^{(LR)}$ and $c_X^{(HR)}$ indicate the number of low- and high-resolution predictor variables per grid node, and $s_{lat}$ and $s_{lon}$ denote the number of grid nodes of the low-resolution array grid in the latitude and longitude directions, as specified in Section 3. Note here that the values of $s_{lat}$ and $s_{lon}$ may equal the maximum values $s_{lat} = 36$ and $s_{lon} = 60$, corresponding to running the model on the full data inputs, but may also be set to smaller values as the convolution operations can adapt to varying input sizes by returning outputs of smaller size, accordingly. Choosing smaller values of $s_{lat}$ and $s_{lon}$ corresponds to running the models on limited sub-domains, which we use for data augmentation, as discussed in Section 5.2. Predictands $y$ are assumed to be of shape $c_Y \times 4s_{lat} \times 3s_{lon}$, with $c_Y$ indicating the number of predictand variables.

While $c_Y = 2$ is fixed for all our models, corresponding to high-resolution wind components $U$ and $V$, $c_X^{(LR)}$ and $c_X^{(HR)}$ vary depending on the predictors supplied to the models, as detailed in Section 6. In particular, some of the models are provided with low-resolution predictors exclusively, whereas other model configurations are informed additionally with high-resolution topography predictors.

The (rectangular) filter kernels are parametrized per convolution layer as arrays of shape $c_{in} \times c_{out} \times k_{lat} \times k_{lon}$, with $c_{in}$ and $c_{out}$ denoting the numbers of input and output features of the layer, and $k_{lat}$ and $k_{lon}$ the spatial extent of the kernel filters in latitude and longitude. Due to the size of the kernel, the number of elements in convolution output arrays differs from that of the input arrays. To compensate for this, suitable replication padding between successive convolution layers is employed to maintain the spatial shape of feature arrays constant throughout the series of convolutions.

In the following, the details of the different model architectures used in our evaluation are described. A schematic summary of all models is provided in Figure 4.

## 4.1 | Linear convolutional network model: LinearCNN

A simple way of mapping the low-resolution data to the high-resolution domain while exploiting the parameter sharing capabilities of CNNs is to learn local linear relationships between predictors and predictands via a linear convolutional model, that is, without nonlinear activation functions. For our experiments, we propose Linear-CNN, an efficient two-layer CNN which is composed of two branches for processing low-resolution and high-resolution inputs separately.

The low-resolution branch is composed of a single standard convolution layer with kernel size $(k_{lat}, k_{lon}) = (5, 5)$, followed by a transposed convolution with kernel size (12, 9) and stride (4, 3). Transpose convolutions can be understood as linear operations which are used to expand the spatial dimension of the input tensors using a linear kernel, which is applied pixel-wise to the inputs of the transpose convolution. A gain in resolution could also be achieved by applying an interpolation scheme, but to let the network learn the proper transformation automatically the transpose convolution is preferred. Striding thereby refers to skipping pixels in the output domain between accumulating successive kernel evaluations and is applied to regulate the difference in resolution between input and output of the transpose convolution (e.g., Dumoulin and Visin, 2016). The architecture of the low-resolution branch of LinearCNN can be thought of as an encoder–decoder scheme. The standard convolution layer transforms the $(5 \times 5)$-pixel input patch into a multi-dimensional $(1 \times 1)$-pixel feature representation, whereas the transpose convolution decodes the features and expands the output to match the resolution of the target domain. Thereby, the dimension of the hidden feature representation can be chosen freely. Settings below $25c_X^{(LR)}$, with $c_X^{(LR)}$ denoting the number of low-resolution predictor variables, correspond to a linear reduction of dimensionality before the decoding step. To maximize flexibility of the model, we choose $25c_X^{(LR)}$ features, thus avoiding implicit constraints on the feature representation. By passing the decoding layer, a $(1 \times 1)$-pixel hidden feature vector is transformed into an output tensor of spatial shape $12 \times 9$ in terms of high-resolution pixels. This output corresponds to a high-resolution estimate of the region, which marks the central $(3 \times 3)$-pixel sub-patch of the $(5 \times 5)$-pixel low-resolution input. Again, the parametrization of the transpose convolution does not constrain the rank of the linear mapping between predictors and predictands. When both convolution kernels are passed across the domain, the high-resolution estimates of neighboring kernel evaluations overlap by 8 and 6 high-resolution pixels in latitude and longitude directions respectively, due to the selected stride values. Effectively, this results in an implicit averaging of predictions from neighboring predictor patches. This is useful to compensate for potential offsets between low-resolution and high-resolution coordinates in latitude and longitude, which may vary across the domain.

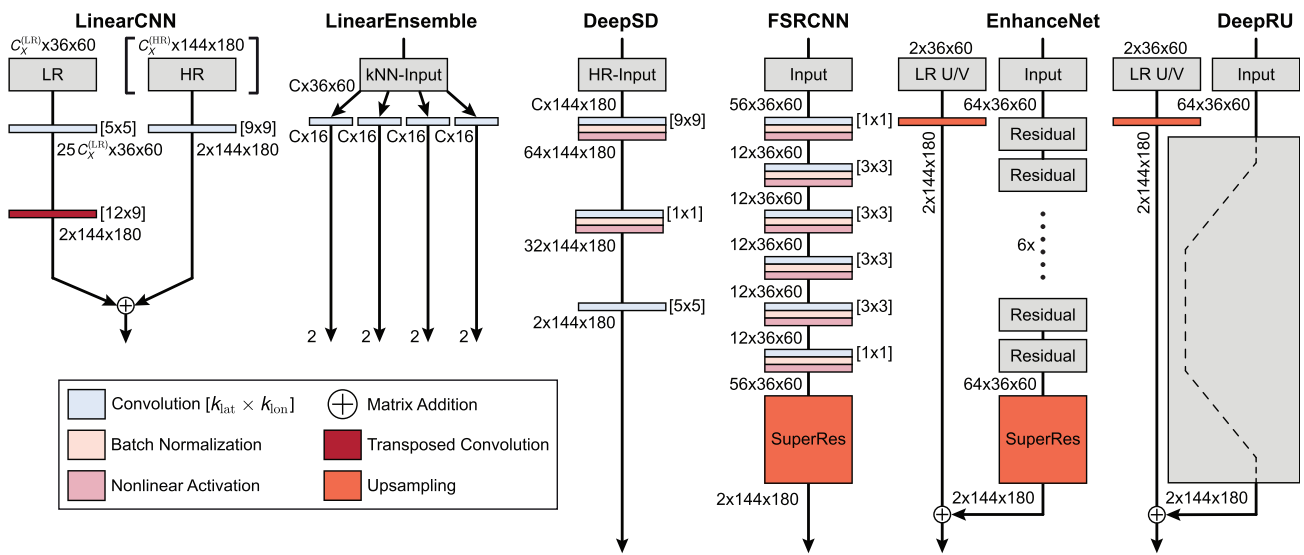

**FIGURE 4**  Schematic of all downscaling models used in this paper. Input sizes of convolutional neural network (CNN) models refer to the final evaluation setting with full domain data. Training was conducted on smaller sub-patches of size $c_X^{(LR)} \times 24 \times 36$ (low resolution) and $c_X^{(HR)} \times 96 \times 108$ (high resolution), as detailed in Section 3

On the high-resolution branch, the predictors are fed into a single standard convolution layer with kernel size (9, 9). The outputs of this layer are directly added to those of the low-resolution transpose convolution. Empirically, we found that models with larger kernel sizes did not improve the performance.

## 4.2 | Simple nonlinear CNN: DeepSD

DeepSD is a simple nonlinear CNN architecture which has been proposed by Vandal *et al.* (2018) for downscaling climate change projections over extended spatial domains. The design of DeepSD builds upon the super-resolution CNN (SRCNN) by Dong *et al.* (2014)—one of the first CNN-based architectures for single-image super-resolution. SRCNN is composed of three convolution layers with rectified-linear activation functions in between, which are used to post-process the result of a bicubic interpolation of the low-resolution image data. Although Vandal *et al.* (2018) proposed composing DeepSD of several instances of stacked SRCNNs for better predictions, we found that for the magnification ratio of 3× in longitude and 4× in latitude a single stage of SRCNN already attains results on a par with those achieved by other SRCNN instances.

In the implementation of DeepSD we follow the design proposed by Dong *et al.* (2014) and Vandal *et al.* (2018). The first layer uses a large kernel size of (9, 9) to transform the input predictor fields into an abstract feature space representation with 64 features, followed by a nonlinear activation. The second layer applies a pixel-wise dimensionality reduction with a convolution of kernel size (1, 1) and 32 output features, and a second nonlinear activation. The final layer applies a convolution with kernel size (5, 5) to transform the features to the target resolution.

Vandal *et al.* (2018) further proposed to inform the model with high-resolution orography to learn the influence of the topography on the inferred climate variables. Hence, we include the high-resolution static orography predictors during training of all our DeepSD models. To match low-resolution and high-resolution predictors, the low-resolution predictors are first interpolated to high-resolution using a bicubic interpolation, and then concatenated to the high-resolution predictors to create a combined input array. A schematic of the high-resolution input (HR-input) block is shown in Figure 5.

## 4.3 | Fast nonlinear CNN: FSRCNN

Beyond previously proposed downscaling models, we also took inspiration from ongoing work in computer vision on image super-resolution. With fast super-resolution CNN (FSRCNN) proposed by Dong *et al.* (2016a; 2016b), we include a direct successor of SRCNN in our comparison.

SRCNN has limitations in computational speed as it operates on a high-resolution interpolant of the original low-resolution image. This leads to an increased amount of floating point operations and requires larger convolution kernel sizes with a large number of trainable parameters to capture spatial features in high resolution. FSRCNN circumvents these problems by applying seven convolution layers to the low-resolution inputs directly and upsampling features to the final target resolution only at the very end. FSRCNN replaces convolution layers with large kernels, that is, (9, 9) or (5, 5), in SRCNN with a sequence of convolutions using smaller kernel sizes of (3, 3) and (1, 1). The smaller-sized convolutions, however, speed up the computation time by a factor similar to the magnification ratio in each dimension and are thus beneficial in terms of inference speed. Dong *et al.* (2016a; 2016b) also proposed an hourglass-shaped network architecture, where the highest number of feature channels is used for the outermost layers, while the channel size of the inner layers is reduced. This design pattern is supposed to avoid costly computations while maintaining prediction quality.

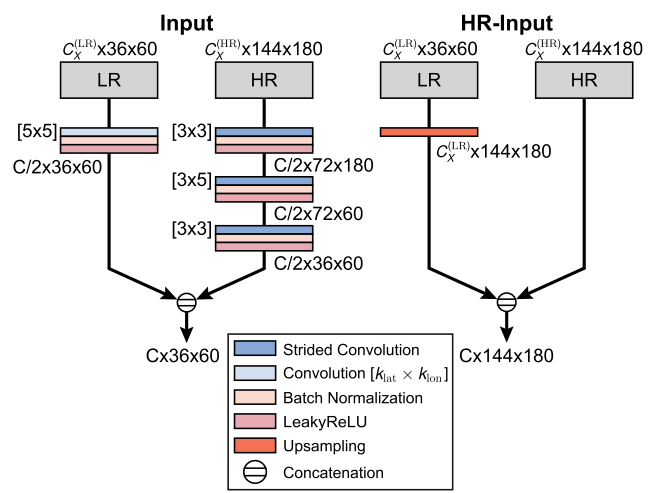

**FIGURE 5** Input blocks used in fast super-resolution convolutional neural network (FSRCNN), EnhanceNet, DeepRU (left) and DeepSD (right)

In our experiments, we slightly adapt the architecture of FSRCNN and split the model into three parts: an input processing stage for primary feature extraction, a feature processing stage and a super-resolution stage for successively increasing the resolution until the target resolution is reached.

The design of the input stage varies depending on the predictors in use. When employing low-resolution predictors exclusively, a single convolution layer of kernel size $(5, 5)$ is used to transform the inputs into a set of 56 spatial feature fields, which coincides with the original design by Dong *et al*. (2016a; 2016b). For model configurations that employ both low-resolution and high-resolution predictors, a combined feature representation in the low-resolution spatial domain is created by applying the input block as depicted in Figure 5. We apply two independent convolution chains to low- and high-resolution predictors separately, and restrict the number of feature channels for both chains to $c^{(LR)} = c^{(HR)} = 28$. While on the low-resolution branch one single convolution with kernel size $(5, 5)$ is used for feature extraction, the high-resolution branch consists of a sequence of strided convolutions with kernel sizes as indicated in Figure 5. This reduces the resolution of the features successively to low-resolution scale. The resulting features are concatenated with the previously computed low-resolution features and supplied to the feature processing stage.

The feature processing stage again reflects the original design choices by Dong *et al*. (2016a; 2016b). In an hourglass-like architecture, a convolution with a $(1, 1)$ kernel is applied to reduce the number of features from 56 channels to 12, which is then followed by a sequence of four convolution layers with kernel size $(3, 3)$, 12 output feature channels, batch normalization and nonlinear activation. The last convolution layer of the processing stage uses a $(1, 1)$ kernel to return to the 56 feature channels.

In the original FSRCNN, the resulting features are used as input for a single transpose convolution with a kernel size of $(9, 9)$ for upsampling. In our experiments, however, we found that this very large transpose convolution can lead to slow training progress and can even prevent training from convergence. Furthermore, Odena *et al*. (2016) have shown that transpose convolutions can introduce checkerboard-like artifacts in the final prediction. To circumvent these problems, the extracted features are fed into a super-resolution block, as sketched in Figure 6, after the final batch normalization and nonlinear activation layer of the feature extraction stage. Hence, we avoid transpose convolutions in our work and, instead, use bilinear upsampling first and apply conventional convolution afterwards to obtain an upsampled result (e.g., Dong *et al*., 2016a; 2016b). In addition, we replace a single upsampling convolution with scaling factor $(4, 3)$ by a sequence of three upsampling blocks with smaller scaling factors of $(2, 1)$, $(1, 3)$ and $(2, 1)$ to obtain the final image in target resolution. The upsampling blocks are composed of bilinear interpolation, convolution layers with kernel size $(3, 3)$, $(3, 5)$ or $(3, 3)$, batch normalization and a nonlinear activation function. Finally, upsampling is followed by an additional convolution layer with batch normalization and nonlinear activation, and a single output convolution without any activation function. Being a nonlinear model, all but the very last convolution layers in FSRCNN are followed by nonlinear activations, which are realized as parametric rectified linear units (PReLU), as proposed by Dong *et al*. (2016a; 2016b).

Note that, in the original FSRCNN architecture, batch normalization was not used. In our experiments, however, we found it beneficial to regularize the feature representations through batch normalization, since the increased depth of our FSRCNN variant may lead to instabilities in training due to, for example, internal covariate shifts (Ioffe and Szegedy, 2015). By applying batch normalization after each convolution, we could successfully stabilize the training process.

## 4.4 | Deep nonlinear CNN: EnhanceNet

Previous work in deep learning (e.g., Timofte *et al*., 2017, and references therein) has shown that increasing network depth can help improve prediction quality and can lead to network architectures which outperform shallow networks. However, deep networks can easily introduce instabilities in the optimization process, which is typically based on backpropagation of gradients. Specifically, training may become inefficient due to vanishing gradients (Glorot and Bengio, 2010), which originate from the accumulation of small parameter gradients in the chain-rule-based estimation of model parameter updates. The sequential algorithm for gradient estimation causes an exponential decay of parameter updates in early layers of the network, and prevents the parameters from changing significantly during training. While batch normalization may help to stabilize network training, vanishing gradients remain an intrinsic problem of deep neural network architectures.

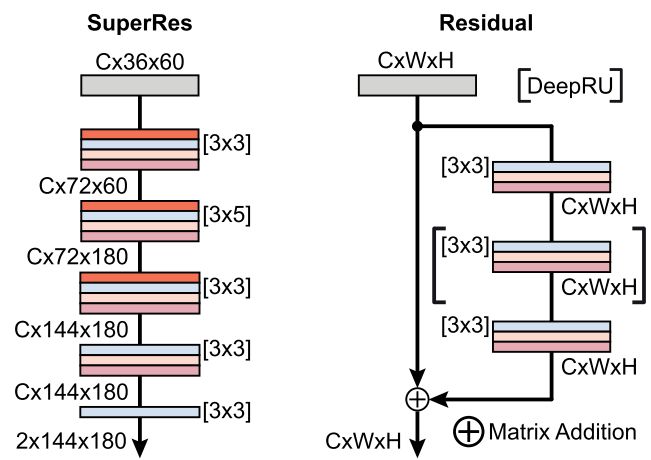

**FIGURE 6** Super-resolution block (left) and residual block (right) for fast super-resolution convolutional neural network (FSRCNN), EnhanceNet and DeepRU

An effective way to address this problem is the integration of so-called short-cut connections. The purpose of these connections is to pass output features of earlier layers directly to a later stage in the network, effectively skipping parameter dependences of intermediate model parts and circumventing the accumulation of small gradients. Two prominent examples are the skip connections used by Srivastava *et al.* (2015) and Ronneberger *et al.* (2015), as well as residual connections proposed by He *et al.* (2016). With skip connections, the output of a previous layer is concatenated with the result of an intermediate layer. An example is given in Figure 7, which is discussed in more detail in Section 4.5. Residual connections are similar to skip connections but, instead of being concatenated, the features before and after intermediate processing are added. This enables the model to learn mappings that are close to identity more directly.

As a deep CNN architecture with residual connections we selected EnhanceNet (Sajjadi *et al.*, 2017), which was originally proposed for image super-resolution. EnhanceNet is composed of an input stage for raw feature extraction, followed by a stack of 20 convolution layers for feature processing and a super-sampling stage (see Figure 6), similar to that of FSRCNN. Residual learning is incorporated into the architecture in two variants. On the one hand, convolutions for feature processing are subdivided into 10 blocks of two layers each, where each block is wrapped by a residual connection. A schematic representation of one of these residual blocks is shown in Figure 6. On the other hand, bicubic interpolation is used to interpolate the low-resolution wind field inputs to target resolution, yielding a baseline estimate for the high-resolution field, which is added to the model output.

For reasons of efficiency, the convolution layers of EnhanceNet use a kernel size of (3, 3). In our experiments, the number of feature channels is set to 64, which is equivalent to the parameters chosen in the original paper by Sajjadi *et al.* (2017). The nonlinear activation functions for EnhanceNet are realized through rectified linear units. Similar to LinearCNN and FSRCNN, we consider network variants with varying settings of low-resolution dynamical predictors, as well as with and without high-resolution topography. Depending on the predictor configuration, either a single convolution layer with kernel size (3, 3) or the input block depicted in Figure 5 is used for primary feature extraction. Since the main focus of our study is on pixel-wise accuracy of the downscaling results, we refrain from using perceptual and adversarial losses that are typically used in super-resolution image tasks (Sajjadi *et al.*, 2017) and instead use pixel-wise losses as discussed in Section 5.3.

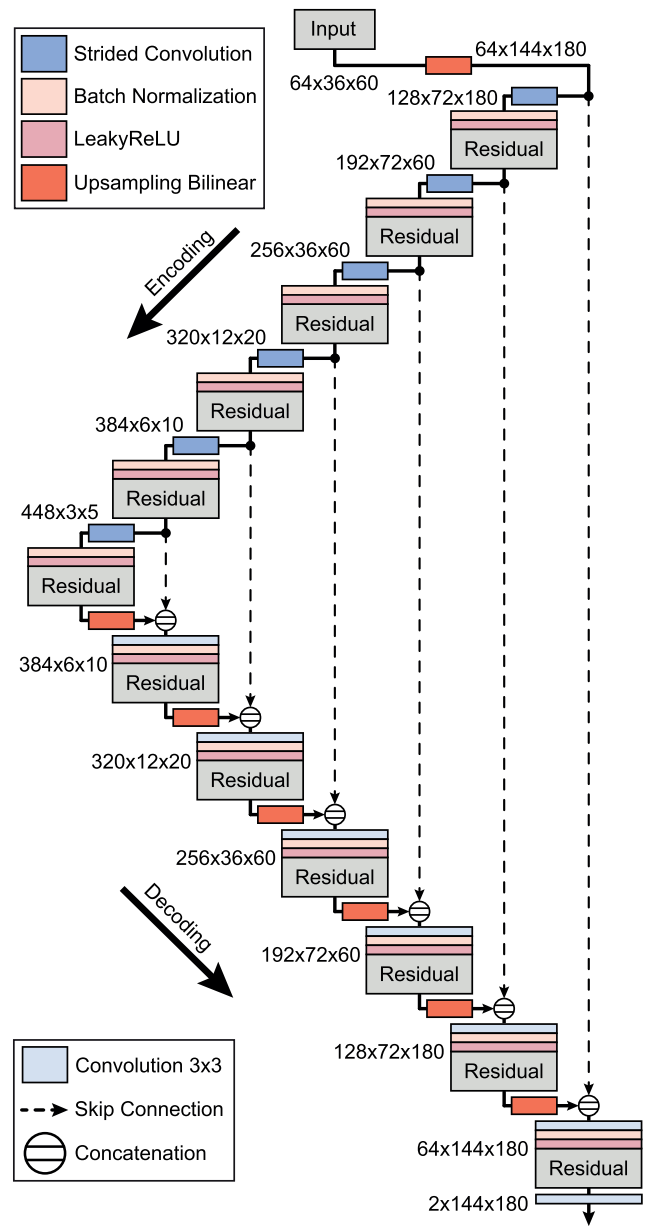

**FIGURE 7**   Schematic of the DeepRU architecture

## 4.5 | DeepRU

Network architectures for super-resolution image generation have been optimized for natural images, which possess properties that are different from those of meteorological simulation results. For instance, natural images typically depict coherent objects, like cars or animals, with well-defined shapes and boundaries. In contrast, meteorological data contain different meteorological variables, which vary smoothly yet less coherently across the domain. Therefore, we expect that more skillful models can be obtained by tailoring model architectures explicitly to meteorological data.

For the present application, we argue that near-surface wind systems result from a complex interplay between a large-scale weather situation, that is, a continental-scale pressure distribution, and boundary-layer processes at finer horizontal scales. The correct treatment of physical processes at varying scales therefore appears as an important aspect in downscaling wind fields on extended spatial domains. This motivates the use of a model architecture that is not restricted to a single resolution scale for feature extraction, but uses different resolution stages to understand the data on multiple scales.

To account for this, we propose to use a U-Net architecture (Ronneberger et al., 2015) with residual connections (He et al., 2016) for downscaling, which we call deep residual U-Net (DeepRU). The U-Net architecture enables an efficient extraction of multi-scale features by design. It consists of two symmetric branches, which are connected by skip connections for simplified information transfer: an encoding branch, on which the data are encoded into an abstract reduced feature representation, and a decoding branch, on which the feature representations are then decoded to reconstruct wind fields at fine-scale target resolution. During the encoding stage, the number of grid points is successively reduced, at the same time increasing the number of feature channels per grid point. In this way, patterns of larger spatial extent can be extracted with small-size convolution kernels. During the decoding stage, the features are super-sampled to a finer scale while reducing the number of feature channels. The skip connections enable a direct information flow between encoding and decoding stages at equal resolutions. By concatenating features from the encoding branch with corresponding features on the decoding branch before further processing, details in the data that could get lost during the compression process can be preserved and localized precisely. In recent work, the U-Net architecture has also been employed for super-resolution tasks (e.g., Hu et al., 2019; Lu and Chen, 2019).

The design of DeepRU is inspired by the results of prior work in image super-resolution (Yang et al., 2019). Starting from the standard U-Net architecture (Ronneberger et al., 2015), we conducted several tests with different U-Net variants to obtain the best model for downscaling. During our studies, we found that making the architecture deeper, that is, increasing the number of resolution levels, led to better training results. The maximum number of levels is limited by the input resolution, since during encoding the input can only be reduced to a tensor of spatial size $1 \times 1$. For downscaling, however, we found that a reduced tensor size of at least $3 \times 5$ in lowest-resolution latent space led to more accurate predictions during patch training and more stable training progress.

While increasing the number of convolution layers with each encoding–decoding stage did not result in better prediction quality, an improvement could be observed when replacing standard convolutions with residual blocks (He et al., 2016). When implementing residual connections across two and even three convolutions at each encoding–decoding stage, we have encountered noticeably improved prediction accuracy.

The reconstruction accuracy could be further improved by interpolating the primary input features after the input block to match the target scale before applying the U-Net model and using skip connections at both high- and low-resolution scale at each encoding–decoding stage. This option gave the most accurate downscaling results between a variety of alternatives that we have tried to process the input. Based on these gained insights, we propose the following architecture for DeepRU.

DeepRU is a six-stage U-Net architecture with both residual and skip connections at every resolution stage (see Figure 7). Similar to FSRCNN and EnhanceNet, we use the input blocks depicted in Figure 5 to transform the inputs to 64 primary low-resolution feature channels. We then super-sample the features using bilinear interpolation, to match the high-resolution grid of size $144 \times 180$. The high-resolution features are then fed into the adapted U-Net architecture. We use strided convolutions to downsample the features during encoding and bilinear interpolation with a successive convolution layer to increase the resolution again during decoding.

At each resolution stage, we apply batch normalization and leaky-ReLU activation before passing features to a residual block, as depicted in Figure 6. The residual blocks, originally proposed by He et al. (2016), have been slightly modified for the downscaling task. We find that extending the original residual block by another convolution layer before the addition operation leads to an increase in flexibility of the residuals, which translates to a better overall model performance.

We implemented skip connections so that a new combined input can be formed by concatenating the features from the encoding stage to the corresponding super-sampled features in the decoding stage. The combined input is then processed by a single convolution layer with batch normalization and leaky-ReLU activation to further reduce the number of feature channels. The reduced features are finally passed to an additional residual block. After the last residual block at the target resolution in the decoding stage, a convolution layer is added to output a set of features which are added to a bicubic interpolant of the low-resolution winds, resulting in the final wind field prediction.

## 4.6 | Localized multi-linear regression model: LinearEnsemble

To enable a comparison of the CNN models with more classical approaches, we also consider a model that is based on standard multilinear regression instead of successive convolutions. Due to simplicity and interpretability, multilinear regression models are frequently used in downscaling and post-processing tasks (e.g., Eccel et al., 2007; Fowler et al., 2007; Gaitan et al., 2014).

For multilinear regression models, Equation (1) can be rewritten in simplified form as

$$y = Wx + b \qquad (2)$$

where $W$ is a $(c_Y d^{(HR)} \times c_X d^{(LR)})$-shaped matrix of weight parameters capturing linear relationships between flattened predictor vectors $x \in \mathbb{R}^{c_X d^{(LR)}}$ and flattened predictand vectors $y \in \mathbb{R}^{c_Y d^{(HR)}}$, and $b \in \mathbb{R}^{c_Y d^{(HR)}}$ is a vector of offset parameters. Again, $c_X$ and $c_Y$ denote the number of predictor and predictand variables per grid node, and $d^{(LR)}$ and $d^{(HR)}$ are the numbers of nodes in the low-resolution and high-resolution domain. Due to the strong increase in the number of trainable parameters with $\mathcal{O}\left(d^{(LR)} d^{(HR)}\right)$ for increasing domain size, typical applications of multilinear downscaling models have been focused on local station data or small spatial domains with limited numbers of grid nodes.

For our comparison, we limit the number of trainable parameters to $\mathcal{O}\left(k \cdot d^{(HR)}\right)$, for some user-defined constant $k \leq d^{(LR)}$. An ensemble of multilinear regression models is trained, where each model uses the $k$-nearest nodes from the low-resolution input to predict the wind components $U$ and $V$ at a single grid node of the high-resolution domain. This corresponds to an induced sparsity pattern on $W$, which allows at most $k \cdot c_X \cdot c_Y \cdot d^{(HR)}$ entries of $W$ to be non-zero.

In contrast to CNNs, we train only two different variants of the model ensemble. In a first step, we use only the low-resolution wind components $U$ and $V$ to inform the model, resulting in a channel number of $c_X = 2$. In a second step, we also add the complementary low-resolution dynamic predictors BLH, FSR and Z500, resulting in a total of $c_X = 5$ predictor channels. Static predictors are not included in the training process, as the resulting contributions in Equation (2) would be indifferent between samples and can thus be incorporated into the offset-vector $b$ without loss of information. The $k$-nearest low-resolution grid nodes are determined based on the standard $L_1$ distance (in latitude–longitude space) to the target node. We empirically determined that neighborhood sizes beyond $k = 16$ did not improve the results significantly in our application.

## 5 | TRAINING METHODOLOGY

The time range of about 3 years that is covered by our data is comparatively short, when set in relation to time scales commonly used to define "climatology." Moreover, temporal correlations between successive samples limit the number of independent examples of weather situations across the domain. This raises the need for efficient data splitting using cross-validation and employing suitable methods to increase the number of training samples. In the following, we shed light on the training methodology and loss functions used in our experiments, and provide details on the optimization process.

## 5.1 | Cross-validation

For all models, including LinearEnsemble, we employ cross-validation with three cycles of model training and validation. In each cycle, we exclude a subset of the data from training. As the data exhibit both short-term temporal correlations on time scales of up to a few days and variations due to seasonality, we decided to pick full consecutive years of data for validation. This minimizes information overlap between training and validation data due to systematic correlations at the beginning and end of validation intervals. Furthermore, it reduces impacts of seasonality on results by averaging model performance over the full seasonal cycle. The excluded validation epochs are chosen pair-wise disjoint and cover the time ranges from June 2016 to May 2017, June 2017 to May 2018 and June 2018 to May 2019, respectively. Each model was trained three times with varying random initializations of the regression parameters in each validation cycle. After convergence, the model with the smallest average validation loss was selected for further evaluation. The performance of the overall model architecture was then assessed by combining the results of the best models of each of the three validation cycles.

## 5.2 | Patch training

To further increase diversity and variance of training samples, we perform CNN training on sub-patches of the full domain. This procedure limits the dimensionality of the model inputs, thus enforcing models to base their predictions on local information and reducing the chance of overfitting to statistical artifacts in the data. Specifically, fitting of potentially non-physical long-distance correlations is efficiently avoided.

From another perspective, patch training is advantageous due to an improved usage of static predictor information in comparison to full domain training. Static predictors remain invariant when training on the full domain and can thus be ignored by the network or be leveraged to establish a network operation mode of local pattern matching, instead of regression. In such a mode, models might learn to associate the invariant topography with preselected local patterns, learned by heart, instead of using the provided dynamic information to regress on.

Confirming our expectations, we found that patch-trained models yield lower training and validation losses compared to models trained on the entire domain. Experiments show that intermediate patch sizes yield the best training results. For very small patch sizes, we observe a decrease in prediction quality, which may be attributed to a loss of context information due to insufficient data supply. These findings may also be related to the concept of the minimum skillful scale of the underlying low-resolution simulation (Benestad *et al.*, 2008), that is, the smallest spatial domain size, for which the low-resolution data provide a sufficient amount of information for the downscaling model to generate skillful predictions.

In our experiments, low-resolution data were processed in patches of size $24 \times 36$ and matched with the corresponding high-resolution patches of size $96 \times 108$. This was found to yield the most accurate full-grid predictions when applied to validation samples. The sub-patches for training were selected randomly for each predictor–predictand data pair and each training step, so that the induced randomness further decreases the chance of overfitting to the training input. Note, however, that patching was applied exclusively during training of the models. For validation and evaluation of model performance, predictions were computed based on the full domain.

## 5.3 | Loss functions

For measuring error magnitude between predictions and high-resolution targets, we consider different deviation measures, which put weight on distinct aspects of reconstruction accuracy. For optimization purposes we consider spatially averaged deviation scores, whereas for further evaluation we consider both average and local deviations.

Given that $\vec{t}_i$ and $\vec{y}_i$ represent the target wind and prediction wind vectors at node $i$, with $1 \leq i \leq d^{(\mathrm{HR})}$ indexing the nodes of the high-resolution grid, we consider in the first place the mean square error (MSE) with

$$\mathrm{MSE}\left(\left\{\vec{t}_i\right\}, \left\{\vec{y}_i\right\}\right) = \left\langle \left|\vec{t}_i - \vec{y}_i\right|^2 \right\rangle_D$$

Here, $\left\{\vec{t}_i\right\}$ and $\left\{\vec{y}_i\right\}$ denote the sets of predictand and prediction vectors throughout the domain at a particular point in time, $|\cdot|$ indicates the standard $L_2$ vector norm and $\langle\cdot\rangle_D$ indicates an average over the spatial domain. The main advantage of MSE is its invariance with respect to rotations of local vector directions, that is, predictand–prediction pairs which differ only by node-wise rotations of wind directions are assigned an identical deviation score.

However, a potential drawback of MSE is that local deviation scores scale quadratically with wind magnitude (the significance of this will ultimately depend on the application). In particular, small-angle deviations in areas of large wind speeds may contribute largely to the overall deviation score, whereas some strong directional deviations, such as opposite wind directions in areas of low wind speed, are hardly taken into account. This problem becomes particularly serious in certain scenarios where slow but strongly variable winds over mountainous areas are accompanied by increased wind speeds over the sea.

A solution to weaken the square dependence effect is to linearize MSE, resulting in the mean absolute error (MAE). Unfortunately, even MAE does not fully overcome the scaling issue and inherits the problems of MSE. Considering angular deviations instead, for instance as measured by cosine dissimilarity, does not provide an alternative either since angle-based deviation measures do not provide the model with information on differences in wind speed magnitude. A potential alternative would be to use a weighted average of the above-mentioned deviation metrics. However, we refrained from using such metrics as this would require an optimization of additional ad hoc hyper-parameters.

An effective solution is to use the standard MSE and reduce spatial inhomogeneity through node-wise standardization of the target predictands. The models then learn to mimic a reduced representation of the non-standard predictands, which can easily be converted back to true scale through an easily invertible linear transformation. As stated in Section 3.5, sample mean and standard deviation are computed from the respective training dataset. For validation and evaluation purposes, we convert back to real-scale target predictands and predictions.

## 5.4 | Implementation and optimization

All models have been realized and evaluated in PyTorch (Paszke *et al.*, 2019). Optimization is performed using the ADAM optimizer (Kingma *et al.*, 2014) with an initial learning rate of $10^{-3}$, which is reduced by a factor of 0.1 whenever the validation loss in terms of MSE does not decay by more than a fraction of $10^{-4}$ over a period of five training epochs. The process is continued until a minimum learning rate of $10^{-6}$ is reached. To guarantee a proper convergence of the models, we train for 150 epochs in each of the three runs per cross-validation cycle, without early stopping. Saturation of training and validation losses was usually achieved after 50–60 epochs, and both training and validation losses showed only minor variations beyond. In particular, we did not observe tendencies of additional overfitting once the models converged.

## 5.5 | Regularization

During training, we employ weight decay with a rate of $10^{-4}$ (Kingma and Welling, 2013). Additionally, nonlinear convolutional models use batch normalization (Ioffe and Szegedy, 2015) after each convolution operation, which we find to accelerate training convergence significantly. For DeepRU, we apply 2D dropout regularization (Srivastava *et al.*, 2014) with a dropout rate of 0.1 after each residual block; that is, succeeding each residual block a fraction of 0.1 of the respective output feature channels is selected randomly and set to zero. Although earlier studies reported performance issues when using batch normalization and dropout regularization in common (see for example Li *et al.*, 2019), we did not encounter any such negative effects.

## 6 | EVALUATION

To compare the different model architectures with respect to downscaling performance, we consider sample-wise deviations between target predictands and model predictions and investigate the extent to which the predictions depend on particular predictors. To shed light on the importance of the choice of predictors, the CNN models are trained with four different predictor configurations, including low-resolution wind fields and orography only, providing supplementary high-resolution orography predictors or additional low-resolution dynamic predictors, or the full set of parameters. The predictor settings are detailed in Table 1 and indicated with letters (A) through (D).

Exceptions from this strategy arise for DeepSD and LinearEnsemble. In the case of DeepSD, we refrain from suppressing the use of high-resolution static predictors in order to stay close to the original implementation, which included high-resolution orography predictors by design. Therefore, for DeepSD, we only consider configurations (B) and (D). For LinearEnsemble we exclude static predictors in both low resolution and high resolution, as by design the model does not take advantage from static predictors (see Section 4.6); we therefore consider only configurations (A) and (C).

## 6.1 | Run-time performance and memory requirements

A general overview of the model performance with respect to the number of trainable parameters, memory consumption and computational time for yearly or daily predictions is provided in Table 2. The time measurements were conducted on the NVIDIA TITAN RTX GPU with 24 GB video memory.

At training time, data for all models except for LinearEnsemble were processed in batches of 30 to 200 samples, depending on the model complexity and memory requirements. During training, a significant amount of the memory consumption is caused by optimization computations which are significantly more complex for deeper model architectures. The measured training time spans the full training period until convergence of the respective model, including prediction time as well as time for loss computation and optimization. In the reference trainings, we considered all dynamic and static predictors at low and high resolution.

LinearEnsemble is exceptional here, as memory limitations arise from the need for rapidly accessible storage of the training data rather than from optimization computations. As the nearest-neighbor positions vary irregularly with spatial position, data selection for LinearEnsemble cannot be realized through efficient array-slicing operations, as is the case for CNNs. Nearest-neighbor indexing has to be performed for all linear models separately and was found to be too slow to be conducted at training time. As a result, data for the

**TABLE 1** Predictor configurations for model trainings with varying combinations of low-resolution (LR) and high-resolution (HR) predictors

| | | | LR | | | | | | | HR | |
|---|---|---|---|---|---|---|---|---|---|---|---|
| | | | Wind | | Dynamic | | | Static | | Static | |
| Configuration | $c_X^{(LR)}$ | $c_X^{(HR)}$ | $U$ | $V$ | Z500 | BLH | FSR | LSM | ALT | LSM | ALT |
| (A) | 4 | 0 | ✓ | ✓ | — | — | — | ✓ | ✓ | — | — |
| (B) | 4 | 2 | ✓ | ✓ | — | — | — | ✓ | ✓ | ✓ | ✓ |
| (C) | 7 | 0 | ✓ | ✓ | ✓ | ✓ | ✓ | ✓ | ✓ | — | — |
| (D) | 7 | 2 | ✓ | ✓ | ✓ | ✓ | ✓ | ✓ | ✓ | ✓ | ✓ |

*Notes:* $c_X^{(LR)}$ and $c_X^{(HR)}$ denote the total number of low-resolution and high-resolution predictor fields supplied to the models.
Abbreviations: ALT, altitude; BLH, boundary layer height; FSR, forecast surface roughness; LSM, land–sea mask; Z500, geopotential height at 500 hPa .

**TABLE 2** Run-time performance statistics for LinearCNN, DeepSD, FSRCNN, EnhanceNet, DeepRU and LinearEnsemble

| Model | TP (k) | MEM (MiB) | TR (hr) | PR (s) | TS (ms) |
|---|---|---|---|---|---|
| LinearCNN | 68.9 | 0.3 | 0.7 | 5.4 | 0.6 |
| DeepSD | 50.6 | 0.2 | 0.9 | 5.8 | 0.7 |
| FSRCNN | 165.3 | 0.6 | 1.9 | 8.0 | 0.9 |
| EnhanceNet | 942.6 | 3.6 | 4.0 | 15.4 | 1.8 |
| DeepRU | 37,113.9 | 142.0 | 13.5 | 82.5 | 9.4 |
| LinearEnsemble | 3,307.4 | 12.6 | 25.8 | 11.8 | 1.4 |

*Notes:* For each model, the columns describe the total number of trainable parameters (TP) in k (thousands), individual memory consumption to store a model (MEM) in MiB, duration of an entire training procedure for a cross-validation run with 8,760 hourly data (TR), prediction time for all 8,760 inputs (PR) and the prediction time for one single time step (TS) in milliseconds.
Abbreviation: FSRCNN, fast super-resolution convolutional neural network.

LinearEnsemble had to be preselected and stored with high redundancy during training. For the full ensemble of 20,416 linear models with 16 nearest neighbors, the 3 year dataset, including all low-resolution dynamic predictors, required the allocation of roughly 137 GiB of memory, which is not feasible to be stored in RAM on a local machine with typically less than 32 GiB available. Hence, the data were outsourced to a separate HDF5 file and streamed from the hard drive during training, which delivers, by a large margin, the highest training time among all trained models. The training times for the remaining models scaled with model complexity, with the highest being for the most complex model—that is, DeepRU.

In contrast to the above, and for reasons of fair comparison, the computational time for model prediction is computed using a batch size of 220 for all networks; note that timings for loss computations and optimization are not included in the measurements. To compute the total time for model predictions, we make use of Python's timer module to measure the plain time required by the model to perform downscaling on all input hours for 1 year, in our case 8,760 hr. As timings are often distorted due to hardware communication and process management, we conducted three measurement runs for all models and averaged the results to obtain the final total prediction time. The time for single hour predictions is represented by the ratio between the total computational time and the total number of inputs. In our study, we experienced that the measured time increased with the model complexity, with highest computational costs for DeepRU.

Regarding the number of trainable parameters, the deeper nonlinear solutions EnhanceNet and DeepRU exhibit a significantly higher number of convolutional layers in comparison to the remaining models and thus require more memory to store the trained parameters. Consequently, the general memory consumption scales with the model complexity (see MEM column in Table 2). Despite the higher consumption of memory for nonlinear models, in particular for DeepRU, we found that they achieved the best overall results in our experiments, which is further discussed in the following sections.

## 6.2 | Quantitative analysis

The statistics of spatially averaged MSE on the validation data are illustrated in Figure 8, confirming that both model architecture and predictor selection have a considerable effect on model performance. The weakest model is LinearCNN, showing the largest overall errors and profiting the least from supplementary predictor information. In particular, the use of high-resolution static predictors, which proved to be useful for all the nonlinear models, appears to have no effect on the performance of LinearCNN. The model appears unsuited to extracting useful correlations between low-resolution predictors and high-resolution wind fields. The reason for this is the restrictive parametrization scheme, which is unsuitable for capturing random offsets and distortions between low- and high-resolution field variables caused by the data padding procedure (see Section 3.4). As the same linear kernels are shared across the entire domain, Linear-CNN is forced to yield a most likely estimate, which, however, is found to be inaccurate for most of the grid nodes and poor regarding spatial detail.

In contrast, LinearEnsemble takes advantage of the local parametrization and achieves considerably better results, comparable with or better than those of the nonlinear models DeepSD and FSRCNN. The gain in performance, however, comes at the expense of a higher tendency of the model to overfit on the training data. In particular, for model variants with a large number of predictors, either due to the use of additional dynamic predictors or larger environment size $k$, one observes severe overfitting. This is visible also in Figure 8, as the maximum reconstruction error of LinearEnsemble models with full predictor set (UV, Dyn, Oro(LR) and UV, Dyn, Oro(LR, HR)) exceeds the maximum error of even LinearCNN. $L_2$ regularization did not improve generalization performance but increased the reconstruction error on both training and test data. For the nonlinear models, in contrast, overfitting could be minimized through weight decay during optimization—having a similar effect as $L_2$ regularization—and dropout regularization.

In agreement with earlier studies by Dong *et al.* (2016a; 2016b), FSRCNN achieves smaller downscaling errors than DeepSD. The quality of the downscaled wind fields, however, is slightly below that of the LinearEnsemble model for all predictor variants under consideration.

Nevertheless, prediction quality can be further improved by considering more complex models. EnhanceNet, which differs from FSRCNN by an increased number of convolution layers and the use of residual connections in combination with bicubic downscaling as additive baseline estimate, is the first model to surpass the performance of LinearEnsemble. Notably, EnhanceNet achieves slightly worse results than LinearEnsemble when omitting the high-resolution orography predictors, but catches up after adding the high-resolution predictors. The same is true for DeepRU, which achieves another reduction of MSE.

Comparing DeepRU and LinearEnsemble directly, we find that DeepRU not only reduces the MSE but can also more effectively take advantage of additional predictors. Whereas LinearEnsemple responds with an increased tendency of overfitting, DeepRU achieves a reduction in deviation score when supplied with high-resolution static and low-resolution dynamic predictors. Specifically, model configuration (D) of DeepRU is the most accurate model in our comparison with an average MSE of around $2.7\,(\mathrm{m{\cdot}s^{-1}})^2$.

## 6.3 | Spatial distribution of prediction errors

To examine the spatial distribution of reconstruction errors, we consider additional angular and magnitude-specific deviation measures, which we average over the sample distribution instead of the spatial domain. Specifically, we consider cosine dissimilarity (CosDis)

$$\text{CosDis}\left(\vec{t}_i, \vec{y}_i\right) = \frac{1}{2}\left(1 - \left\langle \cos\left(\vec{t}_i, \vec{y}_i\right)\right\rangle_X\right)$$

for angular deviations between target predictands and predictions. Systematic deviations in wind speed magnitude are measured in terms of the magnitude difference (MD)

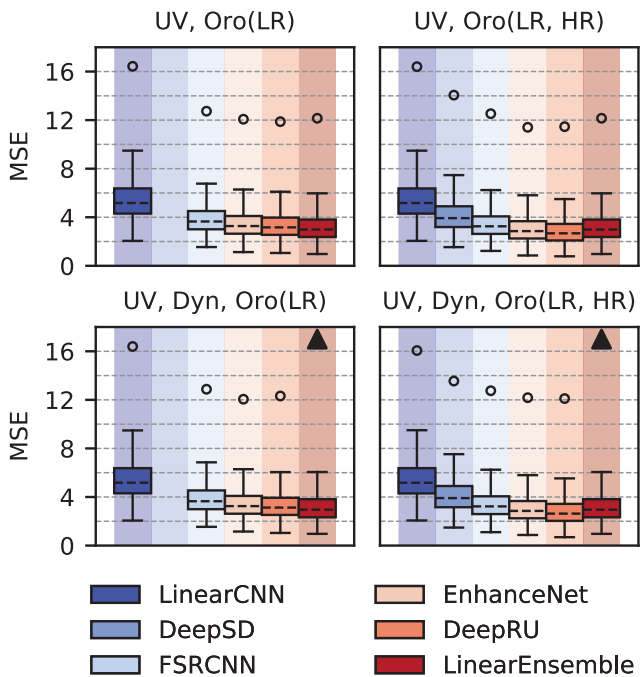

**FIGURE 8** Comparison of validation losses for model variants with varying combinations of input predictors wind components (UV), orography variables altitude (ALT) and land–sea mask (LSM) in low and high resolution (Oro, LR/HR) and supplementary dynamic predictors boundary layer height (BLH), forecast surface roughness (FSR) and Z500 (Dyn). Circles indicate maximum deviation observed on the validation set; black triangles signal maximum reconstruction error beyond the scale of the plot

$$MD\left(\vec{t_i}, \vec{y_i}\right) = \left\langle |\vec{t_i}| - |\vec{y_i}| \right\rangle_X$$

which provides a measure for how much the respective models overestimate or underestimate wind speed magnitudes. In both measures, $\vec{t}_i$ and $\vec{y}_i$ represent snapshots of target and prediction wind vectors at node $i$, and $\langle \cdot \rangle_X$ indicates the sample average over the validation sets of the three cross-validation cycles, respectively.

Figure 9 shows the spatial distribution of magnitude difference and cosine dissimilarity for low-resolution forecasts interpolated bilinearly to the high-resolution grid, as well as outputs of the best-performing DeepRU and LinearEnsemble models relative to the high-resolution forecasts. Regarding the low-resolution simulation, velocities in specific regions near the coasts are not well captured and are mainly underestimated with magnitude shifts greater than 1.0 m·s⁻¹. Angular deviations are more pronounced in mountainous areas. Typical values of cosine dissimilarity range between 0.25 and 0.30, which corresponds to average deviation angles of more than 40°. In the northern part of the Mediterranean Sea, the magnitude difference plot for the low-resolution simulation suggests checkerboard-like artifacts, which, however, are most likely due to a mismatch in spatial resolution and grid structure of low-resolution and high-resolution grids, as well as the use of bilinear interpolation for visualization purposes.

In contrast to the low-resolution simulation, LinearEnsemble tends to underestimate, on average, wind magnitudes at all local grid nodes. We expect that this is mainly caused by an underestimation of extreme winds through LinearEnsemble, which is a common problem of statistical models that are optimized for minimizing MSE losses (e.g., Bishop, 2006). As expected, cosine deviations for LinearEnsemble are much lower than for the low-resolution simulations. However, in areas close to the mountains, LinearEnsemble fails to predict extreme shifts in both magnitude and direction properly, for example due to ridge lines.

DeepRU shows overall better performances with lowest cosine and magnitude differences. Prediction errors exhibit a spatially similar pattern to LinearEnsemble but with generally smaller amplitudes. Furthermore, DeepRU outperforms LinearEnsemble in capturing local variance in wind speed magnitude and directions. As a result, magnitude differences appear less uniform, with overestimation and underestimation in flat areas and near the boundaries, which are caused by imperfect information due to convolution padding. In the Mediterranean Sea, magnitude errors show large-scale wave-like patterns, which especially north of Corsica and east of Sardinia resemble ringing artifacts due to the Gibbs phenomenon (Gibbs, 1898). In turn this relates to the model's spectral representation of topography; issues arise in regions adjacent to where steep slopes meet flat land or sea. In fact the provided topographic height fields contain very similar patterns; sea altitudes look invalid.

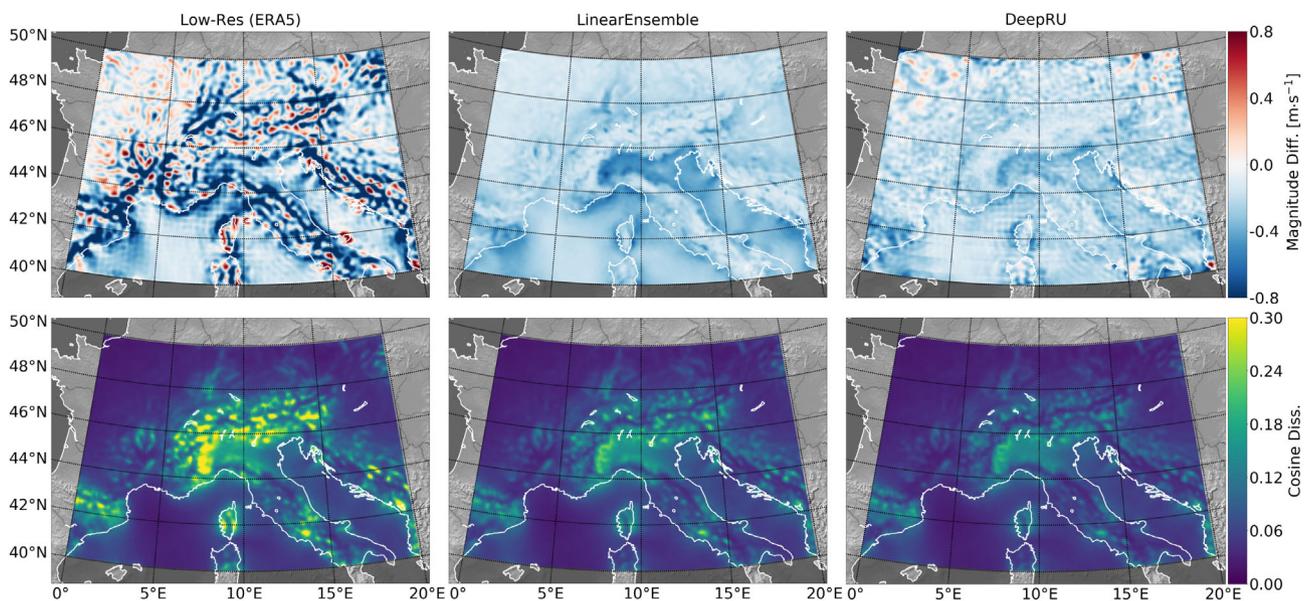

**FIGURE 9**   Mean magnitude difference (top row) and mean cosine deviations (bottom row) between target high-resolution forecast and low-resolution forecast simulation (left), prediction of LinearEnsemble (middle) and DeepRU (right). The average is taken over all three validation years

## 6.4 | Analysis of feature importance

For the model configuration which was trained on the full set of predictors (D), we also investigate the importance of particular predictors according to the method proposed by Breiman (2001). For this, we perturb the model inputs from the validation dataset by randomly shuffling single predictors, and then measure the change in the prediction error that is caused by the perturbation.

Let $X = \{x_1, ..., x_t, ..., x_T\}$ be the (plain) validation dataset for the respective model run, with data samples $x_t = \left(x_t^{(1)} ... x_t^{(p)} ... x_t^{(c_X)}\right)$ containing the predictor variables $x_t^{(p)} \in \mathbb{R}^{s_{lon} \times s_{lat}}$ for $1 \leq p \leq c_X = c_X^{(LR)} + c_X^{(HR)}$. Then, for every predictor $p$ we generate a random permutation $\Pi$ of the sample index set $\{1, ..., t, ..., T\}$, so that the feature-$p$-perturbed dataset $\tilde{X}^{(p)}$ contains samples of the form

$$\tilde{x}_t = \left(x_t^{(1)} ... \, \Phi\!\left(x_{\Pi(t)}^{(p)}\right) ... \, x_t^{(c_X)}\right)$$

Here, $\Phi(\cdot)$ denotes an additional shuffling operation in the spatial domain by decomposing the predictor data into equally sized sub-patches, rearranging the patches randomly and concatenating them again. In our experiments, we fix a patch size of $6 \times 6$. Results for different patch sizes are comparable, though. From the perturbed and non-perturbed predictions $\tilde{y}_t^{(p)}$ and $y_t$, the relative change in prediction error is computed as

$$\rho_t^{(p)} = \frac{\left\langle \mathrm{MSE}\!\left(\tilde{y}_t^{(p)}, y_t^*\right)\right\rangle_{\Pi,\Phi}}{\mathrm{MSE}(y_t, y_t^*)}$$

where $y_t^*$ denotes the ground-truth predictand and $\langle \cdot \rangle_{\Pi,\Phi}$ denotes an average over 10 realizations of $\Pi$ and $\Phi$. Large values of the change ratio $\rho_t^{(p)}$ indicate a stronger impact of predictor $p$ on downscaling accuracy, and thus higher importance of the predictor.

Figure 10 illustrates the sample statistics of $\rho_t^{(p)}$ for the full set of predictors and all downscaling models. In good agreement with expectations, perturbations in the predictor wind components $U$ and $V$ have the largest effects on model performance for all architectures in our comparison, indicating that the models in fact use mainly the information on wind speed and direction for downscaling. The effect of perturbations in the wind components strengthens with increasing model complexity. Reasons for this may lie in the nonlinear structure of the more complex models, which could increase the sensitivity of the predictions to perturbations. Also, as shown in Figure 8, more complex models achieve smaller deviation scores when informed with unperturbed data. A similar increase in prediction error in terms of absolute deviation score therefore yields a larger change ratio for more complex models. This implies that the change ratios $\rho_t^{(p)}$ should be interpreted in a model-specific context.

Assessing the relative importance of the remaining predictors, we find that least information is extracted from FSR, as perturbations in this predictor hardly affect any of the models. As FSR is provided on the same coarse grid resolution as the predictor winds, all the information it provides could already be encapsulated in the winds themselves, so that most models learn to ignore the redundant information. Interestingly, LinearEnsemble is the only model that fits correlations between FSR and high-resolution winds, which may be related to the overfitting problem of the model. Perturbations in BLH also have only a slight impact on prediction performance. This was quite a surprising result, given that this quantity varies considerably over time and given that wind speeds at 100 m can be closely related, especially when BLH values are small.

Z500 is leveraged mainly by the less complex models LinearCNN and DeepSD. Z500 provides information on large-scale weather patterns, and there is a known relationship between its gradients and 500 hPa geostrophic winds, which seems to be recognized most prominently by DeepSD. Nevertheless, direct links between Z500 and 100 m winds tend to be relatively weak, which explains its minor impact on the performance of other models.

## 6.5 | Analysis of reconstructed flow patterns

The quantitative analysis provides high-level abstract information on overall downscaling performance of the models, yet it does not convey detailed information on the ability of the models to reproduce the complex flow patterns that we see in the high-resolution simulation. To investigate this aspect in more detail, we select two example cases, which exhibit strong discrepancies between ERA5 and HRES forecasts, and compare the prediction skills of two different models for these examples. For conciseness, we limit the comparison to outputs of the best-performing nonlinear model DeepRU and the localized linear model LinearEnsemble.

To visualize wind vector fields, we use line integral convolution (LIC), introduced by Cabral and Leedom (1993). To generate a LIC visualization, a randomly sampled white-noise intensity image of user-defined resolution is convolved with a 1D smoothing kernel along streamlines in the vector field. Thus, while LIC generates high correlation between the intensities along the streamlines, different streamlines are emphasized by low-intensity correlation between them. In addition, color mapping is used to encode additional parameters, such as the local vector field magnitude. In contrast to alternative visualizations, such as vector glyphs or streamline plots, LIC provides a global and dense view of the vector field and can avoid occlusion artifacts due to improper glyph size or sparse sub-domains due to improper streamline seeding. A disadvantage of LIC is that there is ambiguity about which of two opposite directions is represented.

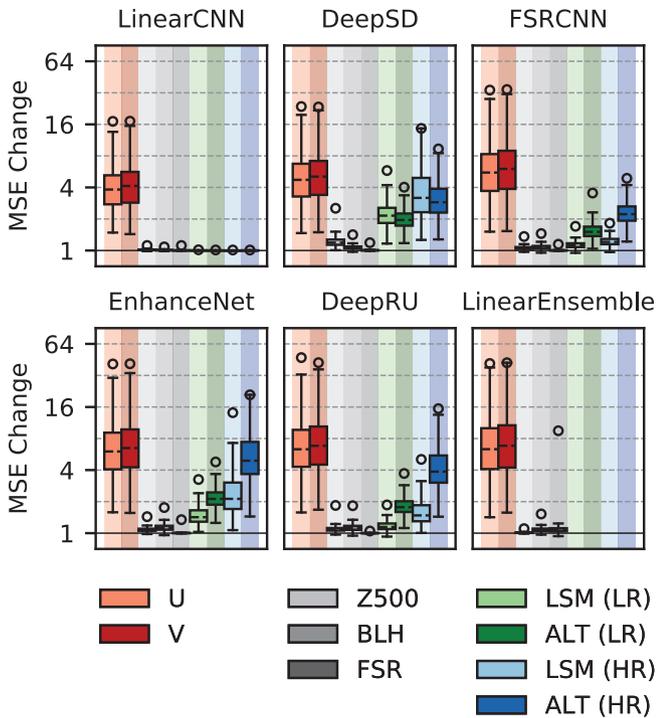

**FIGURE 10**  Relative change in mean square error (MSE) (sample-wise) for different models, when provided with perturbed predictor data. Circles indicate maximum values

The first example is given for lead time October 17, 2017, at 0900 UTC. This case represents a rather anticyclonic scenario with generally low wind speeds, as denoted by the surface charts in Figure 11. Figure 12 shows LIC visualizations of the underlying wind vector fields, obtained from low- and high-resolution forecast simulations. Color coding reflects total wind speed magnitude. Differences in flow patterns indicate that, especially in mountainous regions like the Alps, Apennines (Italy) and Dinaric Alps (Croatia), the low-resolution simulation fails to capture properly the local variability in wind direction and magnitude, which is present in the HRES simulation.

The results of LinearEnsemble and DeepRU are shown in Figures 12c and 12d, respectively. We have highlighted the most important visual differences between the two predictions with rectangles; specific cases are labeled with the letters A–C. In-detail views of the streamlines for all highlighted cases are shown in Figures 13a–c, respectively. Quantitative differences to the HRES simulation are measured in terms of wind direction through local cosine dissimilarity and wind speed through local absolute relative error (ARE)

$$\text{ARE}\left(\vec{t}_i, \vec{y}_i\right) = \frac{|\vec{t}_i| - |\vec{y}_i|}{|\vec{t}_i|}$$

as well as local $L_2$ deviation which combines both aspects. Results for the outputs of the low-resolution simulation and model predictions are depicted in Figure 14.

Based on the quantitative evaluation of all models in Section 6.2, it can be conjectured that both LinearEnsemble and DeepRU reconstruct meaningful downscaling results, with DeepRU leading to overall better prediction quality in scenarios of high inhomogeneities. As seen, for example, in the cases A (Adriatic Sea) and B (Austrian Alps) in Figure 12, LinearEnsemble tends not to reconstruct the flow features when there is a pronounced mismatch in flow patterns between the low-resolution and high-resolution forecast simulations. DeepRU, in contrast, uses both local and global information about the orography, and presumably additional parameters, and is able to replicate the HRES wind fields better. In particular, over the Adriatic Sea (A) the winds are mainly northwesterly, tangential to the coast, and higher magnitudes are more pronounced. LinearEnsemble relies solely on local information in the low-resolution fields and is not able to reconstruct the ground truth faithfully.

In areas of complex surface topography, such as near the Austrian Alps (B), variations in wind speed and direction are usually more pronounced, as wind fields are highly influenced by surface interactions. Here, both models learn a reasonable mapping and are able to handle these cases quite well. According to cosine dissimilarity (Figure 14a), DeepRU performs slightly better than LinearEnsemble in terms of direction predictions. Also, DeepRU is able to replicate extreme transitions in magnitude occurring on small spatial scales better, which results in smaller relative and $L_2$ errors (see Figure 14b,c).

A scenario with generally stronger and rather laminar flow, which exhibits some large differences in wind speed magnitude, is given in (C), where fine-scale mountains slow down winds in eastern France. Since fluctuations in wind direction are small in this area, both models exhibit small errors overall in wind direction. Nonetheless, LinearEnsemble is not really able to account for orography-mediated flow adjustments on small spatial scales, whilst DeepRU can more precisely predict deviations from laminar flow. This is also clearly demonstrated by the absolute relative errors in Figure 14b.

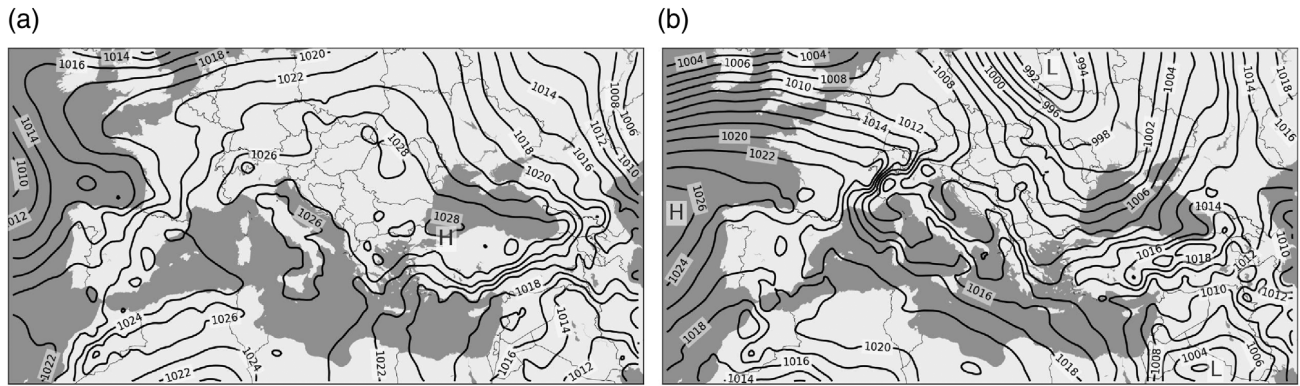

**FIGURE 11** Synoptic charts showing mean sea level pressure (hPa) for 0900 UTC October 17, 2017, and 0100 UTC March 19, 2017

The second example is for March 19, 2017, 0100 UTC. Figure 15 depicts LIC plots of the wind fields for the simulations and predictions similar to Figure 12. As illustrated in Figure 11b, the weather pattern over our domain is mainly dominated by an Alpine lee cyclone, situated between Corsica and northwest Italy. Comparing low-resolution and high-resolution forecast simulations, major parts of the flow are rather laminar with high wind speeds up to 18 m·s$^{-1}$. Contrary to the low-resolution simulation, HRES exhibits sharper changes in magnitude over mountain ridges and mountain edges, and exhibits higher distortions in wind directions over the sea. Two particular cases with differences between forecast simulations and model predictions are highlighted in Figure 15 and are labeled A and B.

In case A, the outputs of both the low-resolution simulation and the LinearEnsemble suggest a rather circular vortex pattern with moderate wind speeds over the Ligurian Sea, between the French Riviera and Corsica. The high-resolution simulation, in contrast, displays a distorted, more elongated flow pattern. DeepRU here elongates the flow around the vortex towards northern Italy and additionally enhances the southerly wind near the western coast of Corsica, which, in summary, better mirrors the predictions of HRES. Case B emphasizes the wind field above northern Italy, where the flow is more inhomogeneous since regions of high wind speeds are interleaved with topographically triggered vortex structures. Here, LinearEnsemble fails to predict as well as DeepRU the sharp magnitude changes seen in HRES along the mountain ridge of the Appenines and near to the three marked lakes.

# 7 | APPLICATIONS IN FORECASTING

As our study sheds lights on the conceptual use of CNNs for downscaling of wind fields, it was not intended that the CNN

architectures proposed here would be used directly in operational forecasting. Indeed the spatial resolutions of our predictor and predictand datasets are not competitive relative to current operational configurations. In Europe, for example, operations nowadays use global models with spatial resolution ~10–20 km, and for shorter leads up to, for example, day 3 use limited area models (LAMs) with resolution ~1–4 km. Nonetheless, our results are sufficiently promising to provide a blueprint for future operational systems that successfully serve the needs of automated and forecaster-based predictions. So how might this work?

To realize this, we envisage first stepping down in scale to use predictor and predictand resolutions of ~5–10 km and ~1–2 km respectively. Regarding the predictors, international modeling centers such as ECMWF will upgrade their global ensembles to this resolution range in the next few years. Regarding the predictand, this is needed only for training and so need not be run operationally in real-time. So one could use, for example, a 1- or 2-year global reanalysis-style dataset, similar to that described by Dueben *et al.* (2020) but created with repeated observation-based initializations. This would deliver worldwide downscaling options, for any region the user selected. An alternative would be to use real-time LAM output for any region for which that was available.

Real-time CNN predictions realized via this route could be used in different ways. Where no LAM coverage exists, predictions could be delivered for short- and medium-range lead times. Where LAMs are available use would focus on the medium range, and if the same LAM were used for training this could nicely provide continuity across the LAM–global temporal boundary.

Another difficulty to address, at least in ECMWF output, is the apparently poor representation of 10 m winds over mountains—the reason we use 100 m winds in this study. This may improve in future, but if not the CNN approach is such that one could use 100 m winds as predictor and 10 m winds as target, if the latter were better represented at 1–2 km resolution—which there is some evidence for, at least for LAMs (Hewson, 2019, Figure 7).

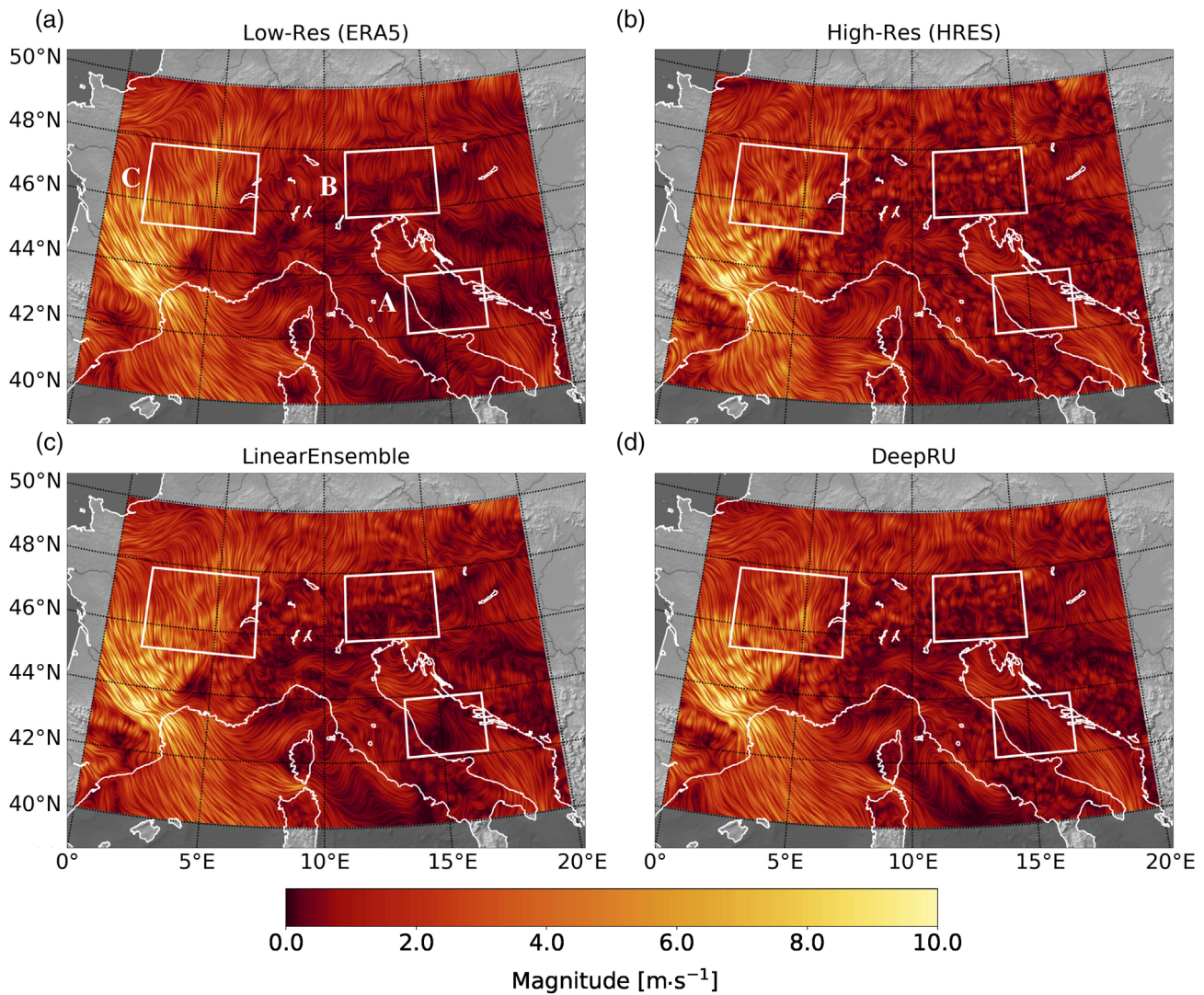

**FIGURE 12**  Wind fields over Europe, as obtained from low-resolution and high-resolution short-range forecast simulations and model predictions for October 17, 2017, 0900 UTC. The top figures show the flow field for (a) the low-resolution and (b) the high-resolution simulation and highlight differences between the two predictions, (c) depicts the predictions of the localized linear model, LinearEnsemble, whilst (d) represents the wind flow predicted by DeepRU. These line integral convolution (LIC) images show the current motion of particle flow produced from the wind field products. The LIC field is colored according to local wind magnitude in m·s⁻¹. Regions with strong differences between predictions are marked by rectangles A, B and C. Errors of LinearEnsemble are MSE = 1.33 (m·s⁻¹)², CosDis = 0.15, and of DeepRU are MSE = 0.88 (m·s⁻¹)², CosDis = 0.097

There are numerous application areas that need better, locally refined wind speed predictions. Renewable energy is clearly one. Others include local pollutant dispersal, coastal and open water shipping, rig operations, leisure activities such as sailing, aviation, the construction industry and warnings in general. Applications for which mean speed predictions are important across the full speed range, such as renewables, will potentially benefit most. For applications with a focus on extremes more investigation will be needed; the training period may not be sufficient. Predictions may be systematically too weak, or become unstable. For very hazardous but less rare gap-flow phenomena we can be more optimistic, however. Here we expect the CNN predictions to deliver major benefits for users compared to raw model output.

Society requires not only predictions of mean wind speeds, but also forecasts of gusts, particularly extreme gusts, because of the dangers posed to life and infrastructure. Gusts have not been directly explored in this study. One might be able to convert mean speeds into reasonable gust forecasts using empirically defined gust-to-mean relationships (see for example Ashcroft, 1994), developed for different land surface types, although for cyclone-related gusts, which tend to be the major wind-related hazard in the vicinity of storm tracks (e.g., in northern and western Europe), caution is needed. Low-level stability, and destabilization mechanisms, as outlined in Hewson and Neu (2015), are of paramount importance for determining the strengths of phenomena such as the cold jet, warm jet and sting jet (see also Browning, 2004). In that context it is curious that the BLH parameter used in our study, which relates directly to stability, did not add much predictive value for the CNNs. Our use of a region that is relatively remote from storm tracks may explain this.

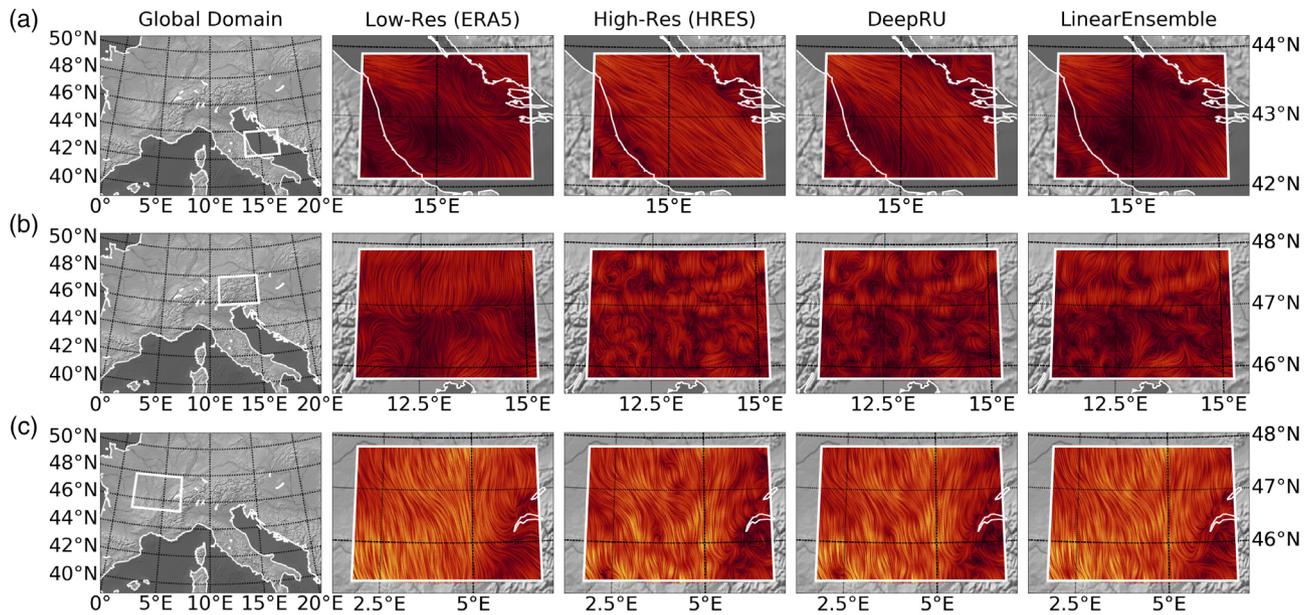

**FIGURE 13**  Example flow patterns on 0900 UTC October 17, 2017, as obtained from low-resolution and high-resolution short-range forecast simulations, and predictions of LinearEnsemble and DeepRU, visualized as line integral convolution (LIC) plots. The location of the regions within the data domain is marked on a global map on the left for each case. (a) The flow field outputs in a region between Italy and Croatia over the Adriatic Sea, (b) the flow over the Austrian Alps with low-speed winds and large directional variations, and (c) the wind flow of areas near central France

It is important to reiterate that airflow, and thus winds, can be very scale dependent. On meter scales speeds around city buildings vary dramatically, whilst on a lake the behavior of a yacht can be influenced by clumps of bushes nearby. Indeed scale dependence is more acute than it is for other parameters, such as rainfall and temperature. Thus model resolution increases bring with them more and more application areas for forecasts, particularly for regions that are topographically and/or physically complex. In turn this brings sustainability, whereby the method outlined in this paper, and variants of it, can find utility for the foreseeable future as numerical weather prediction models continue to evolve.

## 8  |  DISCUSSION AND OUTLOOK

Driven by fast developments in computer science, applications of data-driven machine learning methods in a meteorological context are attracting increasing interest. In the current study, we have investigated the use of CNNs for learning-based downscaling of wind fields. However, the sheer volume of potential design choices which could impact model performance tends to preclude a complete understanding of reasons for the performance of particular model architectures. Therefore, we have selected a set of design patterns based on the experience of what model types have worked well on similar tasks. Our proposed final architecture marks the preliminary endpoint of an iterative process of repeated model training, evaluation and architectural refinement, and was found to achieve the most promising performance in our application. It is clear, on the other hand, that even with only a limited range of design patterns the computational cost of training a large number of CNNs rules out a complete and direct comparison of model architectures. Thus, given the ever-increasing number of studies in data science and machine learning, it can be expected that alternative architectures can be found that achieve similar or superior downscaling accuracy, ideally with reduced model complexity.

Our study has shown that the prediction accuracy of a linear ensemble model is higher than what can be achieved with shallow nonlinear CNN architectures. In particular, for simplistic nonlinear models with only a few convolution layers, it seems that the nonlinearity even hinders performance. We attribute this to distortion of the wind field information by the nonlinear activations on its way through the network, which prevents the model from benefitting from simple mapping schemes, such as for example interpolation kernels. Thus the use of overly simplistic and shallow nonlinear models may be one reason why earlier studies found hardly any additional value in applying neural-network-based machine learning methods (e.g., Eccel *et al.*, 2007; Vandal *et al.*, 2019).

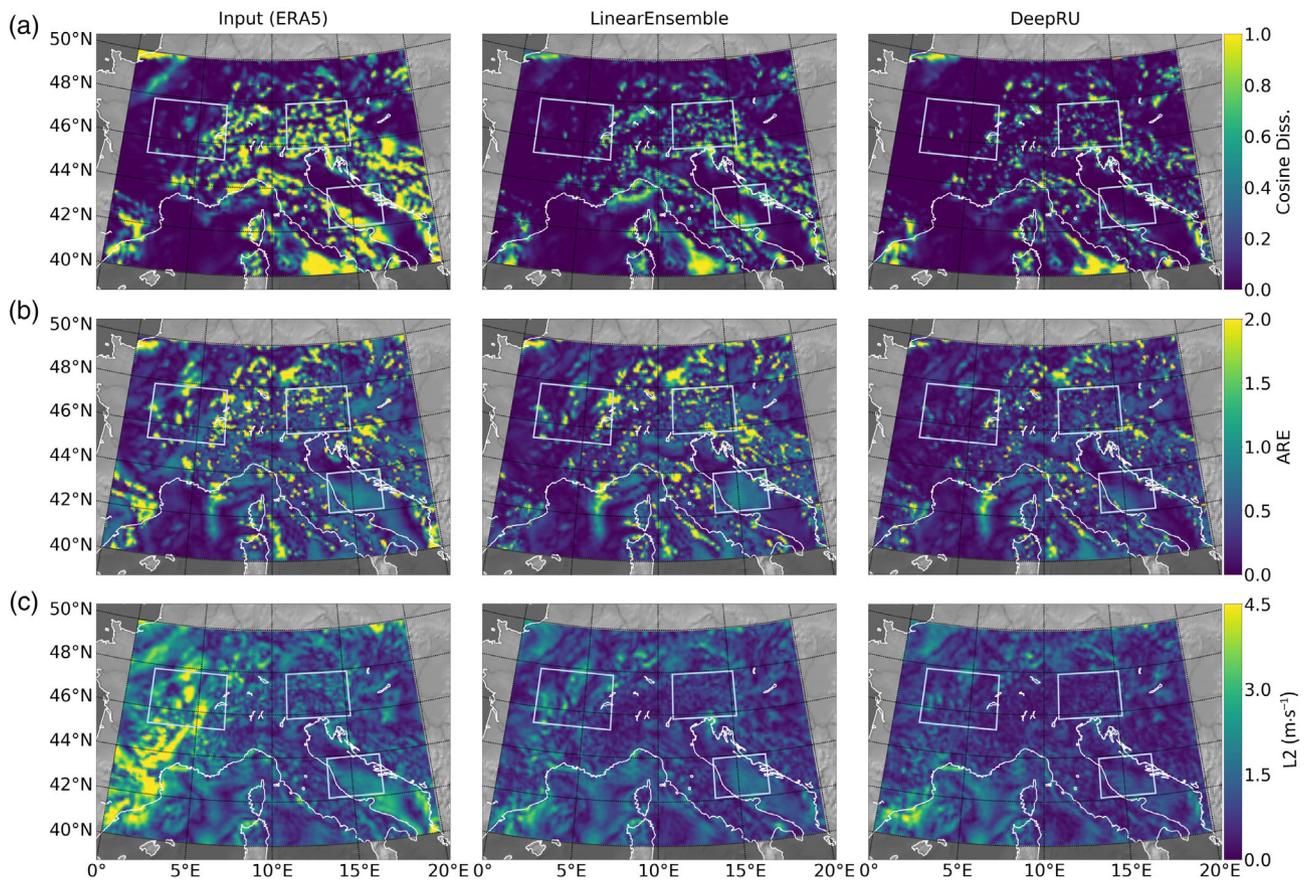

**FIGURE 14**   Visualization of spatial deviations of the low-resolution simulation, LinearEnsemble, and DeepRU wind predictions compared with the output of the high-resolution simulation shown in Figure 12. Here, the deviations are (a) cosine dissimilarity, (b) absolute relative error and (c) $L_2$ norm

Deeper nonlinear CNNs, on the other hand, are able to compete with the prediction quality of the linear ensemble model and even show superior results when incorporating an increasing number of predictors and high-resolution topographic information. In particular, we identified EnhanceNet, previously proposed for single-image super-resolution, as a deep CNN that achieves this. As seen in Figure 9, EnhanceNet exhibits a clear increase in prediction quality with additional parameters while LinearEnsemble is unable to make use of this information and tends to overfit on the training data, finally with an overall slightly inferior prediction performance. EnhanceNet thus appears more flexible and minimizes the need for incorporating prior knowledge and manual selection of suitable predictor variables. Instead one can select *candidate* predictor variables and refine those later based on an analysis such as is shown in Figure 10.

With DeepRU, we propose a novel deep residual U-Net architecture, which outperforms both the linear model and EnhanceNet in terms of reconstruction accuracy. The major advantage of DeepRU lies in its ability to process features at different spatial scales. This is particularly useful for downscaling of wind fields, where local wind systems have to be consistent with large-scale flow patterns. Although we still observe some deviations between high-resolution model predictions and native high-resolution forecast simulations, we are confident that CNNs can provide promising downscaling results and add more value to downscaling than linear models at a reasonable computational cost.

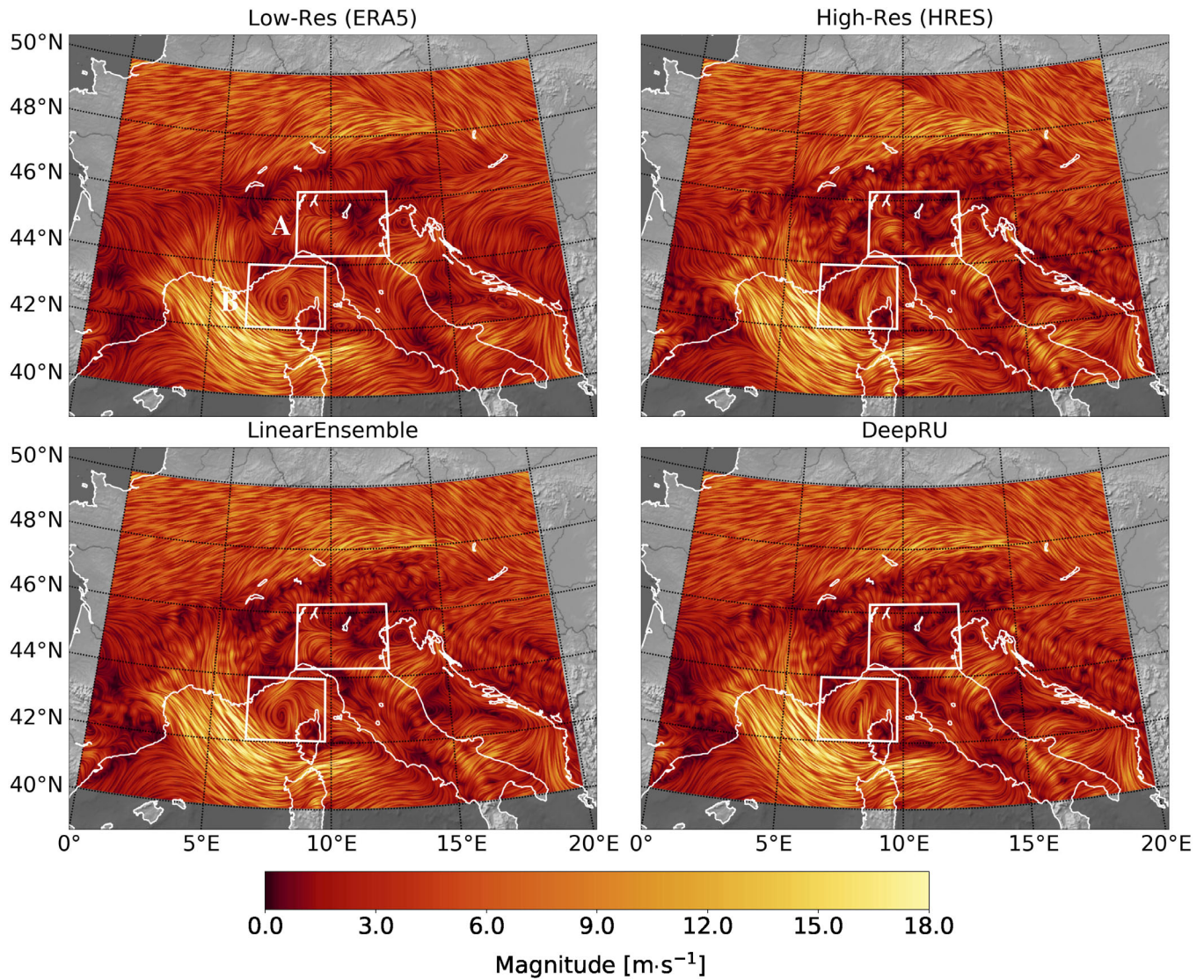

**FIGURE 15**  Wind fields over Europe, as obtained from low-resolution and high-resolution short-range forecast simulations and model predictions for March 19, 2017, 0100 UTC, similar to Figure 12. Color coding indicates the local wind velocity

Our study suggests that deep CNN approaches are particularly effective for downscaling with high magnification ratios on large spatial domains. In this setting, the use of classical models becomes computationally inefficient, and linear link functions between predictor variables and predictands become insufficient to account for non-trivial variability in the local flow, for example due to pronounced flow distortion around obstacles. We found that deep CNNs are better suited to replicating this variance, especially in mountainous areas or over the sea near to coasts, and expect that the same holds true also at finer spatial scales.

Important aspects that need to be further examined in the future are model verification and generalizability. In our study, we have trained CNNs on downscaling tasks using wind fields over a particular spatial domain, that is, the predictive skills of the resulting networks are specific to this concrete setting. Low-level winds were selected as a variable assumed to be particularly appropriate for this type of methodology, because their structure in the vicinity of coasts and complex topography is very much determined by those physical features together with the broader scale flow patterns delivered by ERA5. Downscaling of some other climate variables will require different modeling approaches, because physically the problem can be very different. Each variable, and suitability thereto, must be considered individually. For accumulated precipitation, for example, the range of possible outcomes at high resolution, for a given low-resolution representation, might be limited for one type of precipitation (e.g., orographically enhanced) but considerable for another (e.g., convective) and therefore precipitation downscaling lends itself to a completely different and innately probabilistic approach (e.g., Hewson and Pillosu, 2020). Even so, there will be climate variables other than low-level winds for which, given suitable predictor variables, the model architectures proposed in this paper can serve as flexible feature extractors and yield skillful downscaling results.

An alternative notion of generalizability refers to applying readily trained network models to predictor data which formally depict the same climate variables as the data used during training but deviate from the training set with respect to certain properties, such as geographic reference domain or applied simulation model. In such cases, we expect a poorer performance. Specifically, we have applied our networks to wind field data over a region in North America, covering parts of the west coast of the northern United States and Canada, as well as the Rocky Mountains and parts of the interior plains. Although the data were generated with the same simulation procedures as for the original training data over Europe, we observed a drop in performance of about 70% in MAE and 90% in cosine dissimilarity. When training directly on the data from North America, however, similar reconstruction quality as reported in this paper could be achieved also on the other domain. Our findings indicate that the generalizability of our CNN-based downscaling approaches should be assessed carefully. One possible workaround for future applications could be accepting the lack of portability of the models and training many different networks, each of which is specialized and validated within its particular scope. Additionally, though, it seems promising to examine further how networks can learn to model concurrently the relationships occurring between meteorological variables over a variety of different domains and data sources. Our results suggest that the apparent lack of generalizability is not due to insufficient flexibility of the models, which is in line with earlier work on generalizability of deep learning models (e.g., Zhang *et al.*, 2016). Specifically, our models can be taught to achieve high reconstruction scores over both domains, North America and Europe, when data from both regions are seen during training. The main focus should thus be on increasing the data efficiency of the models to facilitate generalization, for example by incorporating prior physical knowledge concerning recurring atmospheric processes into model design or training regularization.

What we have neglected so far in this paper is the temporal dimension of the data, which can probably be used to understand the model predictions better and further to improve their performance. In preliminary research, we have assessed how temporal correlations are reflected in the model's predictions and found that temporal correlation between model predictions and target wind fields yields information complementary to that conveyed by MSE measurements. In particular, we found that, according to temporal correlation, our models exhibit highest uncertainty over mountains while MSE deviation is largest over the sea. In the present experimental setting, however, the role of the temporal dimension is more similar to that of a sample index, instead of a temporal coordinate, which parametrizes the time evolution of physical processes. Specifically, training of our proposed models has focused on purely spatial correlations on a single-time-step level and temporal coherence between predictions has not been enforced. Consequently, it would be interesting in the future to design neural network models which consider the temporal correlation of wind vector fields across multiple time steps and analyze the models in terms of predictability. This would require the definition of a suitable and interpretative temporal correlation measure for vector-valued inputs which, in our opinion, appears to be a non-trivial task. For instance, we have found that the temporal average of the scalar product between mean-centered predictor and predictand wind vectors, as a standard correlation measure, strongly resembles cosine deviations, which we attribute to the strong relationship between the scalar product and the definition of the cosine deviation. Another option would be to examine local coordinate-wise temporal correlations between the scalar wind components, which, however, would require the selection of reference directions for computing these correlations. The best candidates for these may not be known beforehand and presumably depend on the local surface topography. Beyond coordinate-wise correlations, full correlation matrices might be necessary to examine existing cross-correlations between wind components in a complete and principled way.

Furthermore, including temporal information also into the process of model building (e.g., using long short-term memory (Hochreiter and Schmidhuber, 1997), gated recurrent units (Cho *et al.*, 2014) or related temporal neural network building blocks) or model training (e.g., using optimization objectives, which enforce temporal coherence) could be an interesting direction for future research. To be convincing to an end-user, one ultimately wants the time-series coherence in predictions for given sites to be comparable to time-series coherence in the training data, and therefore devoid of odd jumps except those that are physically realistic—for example due to passage of a front. The current time-independent approach is good in that it might help preserve frontal passage wind-shifts at points, but on the other hand this may possibly be at the expense of other unexplainable temporal shifts in wind velocity.

Another important question for future research, which directly follows on from these ideas, is how to account for the spherical domain geometry in CNN-based downscaling. While data padding was found to be well suited for reshaping irregular grids on domains of up to a few thousand kilometers of horizontal extent, increasing domain size even further may lead to distortion artifacts due to disregard of the spherical geometry of the Earth's surface. The same is true for interpolation-based resampling methods, where the horizontal spacing of the sampling points varies with latitude, limiting data resolution close to the equator and enforcing data redundancy closer to the poles. Furthermore, an inappropriate treatment of domain geometries might become a serious problem, especially for models which are supposed to work on multiple domains. The use of more appropriate convolutional model architectures, like spherical CNNs for unstructured grids (e.g., Jiang *et al.*, 2019) or geometric deep learning approaches in general (e.g., Bronstein *et al.*, 2017), may help to overcome such limitations, thus increasing physical plausibility and data efficiency of the models.

From the exciting perspective of real-time application, one would ideally want to step down in scale and apply the results of this proof-of-concept study in a finer resolution setting. We envisage that operational real-time forecast runs—single deterministic and/or ensemble—could be downscaled in real-time to 1–2 km, over any preselected domains, for customer applications. This could be activated on a central cloud-type platform or locally by customers to meet their own needs. Given the small number of low-resolution predictors, data transfer requirements for the second option would be minimal, compared to say the task of transferring 4D (full-atmosphere) fields for many variables.

At such very high target resolutions, particularly if a high multiplier were used, the correct treatment of ambiguity in the data becomes increasingly important, since the same coarse-scale flow pattern may correspond to multiple fine-scale realizations. Similar to stochastic weather generators, generative CNN models like variational auto-encoders (Kingma and Welling, 2013) or convolutional generative adversarial networks (e.g., Goodfellow *et al.*, 2014; Radford *et al.*, 2015) may provide promising approaches for building flexible models for ensemble-based probabilistic downscaling. Moreover, if the low-resolution feed were based on ensemble data itself, one could then generate a super-ensemble (i.e., ensemble of ensembles) to provide the final smooth-format probabilistic output for users.

## 9 | CONCLUSION

In this study, we have analyzed convolutional neural networks (CNNs) for downscaling of wind fields on extended spatial domains. By going from a simple linear CNN to deeper and more elaborate nonlinear models, we have investigated how the network complexity affects downscaling performance. We have further compared the performance of different CNNs to that of an ensemble of localized linear regression models.

We have shown that deeper and more complex network models are able to discover skillful mappings by exploiting nonlinear correlations for modeling the relationship between low- and high-resolution fields. Specifically, we found that all nonlinear models in our study take advantage of additional high-resolution static predictor data, such as information on local orography. In comparison, the use of three pre-defined low-resolution dynamic predictors gave only minor improvements.

Building upon the results of our study, we have envisioned a number of possible further research directions, like inclusion of temporal information into the training process, or examination of generative neural network models for probabilistic downscaling. We firmly believe that the demonstrated performance of CNNs for downscaling tasks should motivate further research towards the use of such architectures for predictive tasks.

## ACKNOWLEDGEMENTS
This research has been done within the subprojects B5 and A7 of the Transregional Collaborative Research Center SFB/TRR 165 Waves to Weather funded by the German Research Foundation (DFG). We thank all reviewers for their constructive criticism and valuable comments. Open access funding enabled and organized by Projekt DEAL.

## ORCID
*Kevin Höhlein*  https://orcid.org/0000-0002-4483-8388
*Michael Kern*  https://orcid.org/0000-0002-8060-3367
*Timothy Hewson*  https://orcid.org/0000-0002-3266-8828
*Rüdiger Westermann*  https://orcid.org/0000-0002-3394-0731